# Distinguishing two-component anomalous Hall effect from topological Hall effect

*Lixuan Tai[1†], Bingqian Dai[1†], Jie Li[2†], Hanshen Huang[1], Su Kong Chong[1], Kin Wong[1], Huairuo Zhang[3,4], Peng Zhang[1], Peng Deng[1], Christopher Eckberg[1,5-7], Gang Qiu[1], Haoran He[1], Di Wu[1], Shijie Xu[1,8], Albert V. Davydov[4], Ruqian Wu[2\*], and Kang L. Wang[1\*]*

1. Department of Electrical and Computer Engineering, University of California, Los Angeles, California 90095, United States

2. Department of Physics and Astronomy, University of California, Irvine, California 92697, United States

3. Theiss Research, Inc., La Jolla, California 92037, United States

4. Materials Science and Engineering Division, National Institute of Standards and Technology (NIST), Gaithersburg, Maryland 20899, United States

5. Fibertek Inc., Herndon, VA 20171

6. US Army Research Laboratory, Adelphi, MD 20783

7. US Army Research Laboratory, Playa Vista, CA 90094

8. Shanghai Key Laboratory of Special Artificial Microstructure and Pohl Institute of Solid State Physics and School of Physics Science and Engineering, Tongji University, Shanghai 200092, China

ABSTRACT: In transport, the topological Hall effect (THE) presents itself as non-monotonic features (or humps and dips) in the Hall signal and is widely interpreted as a sign of chiral spin textures, like magnetic skyrmions. However, when anomalous Hall effect (AHE) is also present, the co-existence of two AHEs could give rise to similar artifacts, making it difficult to distinguish between genuine THE with AHE and two-component AHE. Here we confirm genuine THE with AHE by means of transport and magneto-optical Kerr effect (MOKE) microscopy, in which magnetic skyrmions are directly observed, and find that genuine THE occurs in the transition region of the AHE. In sharp contrast, the artifact "THE", or two-component AHE occurs well beyond the saturation of the "AHE component" (under the false assumption of THE+AHE). Furthermore, we distinguish artifact "THE" from genuine THE by three methods: 1. Minor loops, 2. Temperature dependence, 3. Gate dependence. Minor loops of genuine THE with AHE are always within the full loop, while minor loops of the artifact "THE" may reveal a single loop that cannot fit into the "AHE component". Besides, the temperature or gate dependence of the artifact "THE" may also be accompanied by a polarity change of the "AHE component", as the non-monotonic features vanish, while the temperature dependence of genuine THE with AHE reveals no such change. Our work may help future researchers to exercise cautions and use these methods to examine carefully in order to ascertain genuine THE.

The anomalous Hall effect (AHE) is a well-known phenomenon in ferromagnets. In a homogeneous ferromagnet, the AHE is linearly proportional to the magnetization normal to the plane, $M_z(H_z)$

$$R_{AHE}(H_z) = R_S M_z(H_z) \qquad (1)$$

where $R_{AHE}(H_z)$ is the AHE resistance. This makes $R_{AHE}$ a monotonically increasing (or decreasing) function of the applied perpendicular field $H_z$ up to magnetic saturation at high field.[1] However, sometimes pronounced nonmonotonic anomalies, or humps and dips in the Hall resistance can be observed and are usually taken as the topological Hall effect (THE).[2]

The origin of the THE is distinct from that of the AHE. The AHE, as a combined effect of exchange and spin-orbit interactions, has both extrinsic contributions including skew scattering[3,4] and side jump[5,6] and the intrinsic contribution from the Berry curvature in the momentum space.[7] By contrast, the THE is from the extra real-space Berry phase obtained by an electron moving in topologically nontrivial (or chiral) spin textures, which effectively work as an extra fictitious magnetic field and deflect the electrons perpendicularly to the current direction without the need of spin-orbit coupling.[2,8] Among all chiral spin textures that might contribute to the THE, the most notable ones are magnetic skyrmions.

Magnetic skyrmions are swirling topological spin structures with particle-like properties and have the prospect of working as bits in memories and logic devices with the advantages of high-density, non-volatility and low power consumption. The topological properties of magnetic skyrmions are characterized by its topological charge, which captures the number of times the unit vector of the magnetization $\mathbf{n}(\mathbf{r})$ winds across the entire unit sphere in real space, and is given by [9,10]

$$Q_{sk} = \frac{1}{4\pi} \int d^2 r \, \mathbf{n} \cdot \left( \frac{\partial \mathbf{n}}{\partial x} \times \frac{\partial \mathbf{n}}{\partial y} \right) \qquad (2)$$

Magnetic skyrmions can be stabilized by the anti-symmetric Dzyaloshinskii-Moriya interactions (DMI),[11,12] dipole-dipole interactions[13] or frustrated exchange interactions.[14] Predicted in magnetic metals lacking inversion symmetry,[15] magnetic skyrmions were first observed in experiments by means of neutron scattering,[16] and later by Lorentz transmission electron microscopy,[17,18] magnetic force microscopy,[19] magneto-optical Kerr effect microscopy,[20–23] resonant soft X-ray diffraction,[24] photoemission electron microscopy,[25] etc.

The THE is generally accepted as a hallmark of magnetic skyrmions and found to be linearly proportional to 2D density of total topological charges from magnetic skyrmions,[9,10] since the THE in the A phase of MnSi[26,27] was found to be consistent with the direct observation of skyrmions by neutron scattering.[16] Such consistency was also found in later reports of FeGe,[28,29] B20-type MnSi,[30] Ir/Fe/Co/Pt multilayers,[31–33] and several other materials,[34–36] and single skyrmions can be detected directly by Hall voltage.[37] Therefore, a large number of experimental studies to date used the THE alone as the sufficient evidence for chiral spin textures, including magnetic skyrmions.[38,39,48–57,40–47] However, in the presence of AHE, the two-component AHE, i.e. the co-existence of two AHE loops, may also give rise to a non-monotonic shape that is similar to genuine THE with AHE, making it difficult to distinguish between the two.[58–61] Distinguishing the genuine THE from such artifacts would generally require the direct observation of chiral spin textures (including magnetic skyrmions) in real or momentum space, by resorting to advanced techniques like Lorentz transmission electron microscopy or magnetic force microscopy. To date, the transport studies alone for identifying magnetic skyrmions could be unreliable.

It is thus highly desirable to find the unique features that can distinguish two-component AHE from THE directly in transport.

In this Article, we use magnetotransport and magneto-optical Kerr effect (MOKE) microscopy simultaneously to confirm magnetic-skyrmion-induced genuine topological Hall effect (THE) + anomalous Hall effect (AHE). We compare this with the two-component AHE (or artifact "THE") and summarize their unique features, and further develop three methods to distinguish their differences: 1. Minor loops, 2. Temperature dependence, and 3. Gate dependence. Our test materials for genuine THE with AHE are thin films of Ta(5)/CoFeB(0.9)/Ir(0.03-0.15)/MgO(2)/Al$_2$O$_3$(5) grown by magnetron sputtering (numbers in parentheses represent thickness in nm), while the ones for two-component AHE are magnetic topological insulators, MnBi$_2$Te$_4$, grown by molecular beam epitaxy with the secondary phase MnTe$_2$ contributing to the extra AHE component. Because our methods can be quickly and easily checked in transport, we believe that they are useful for distinguishing the artifacts from genuine THE induced by magnetic skyrmions.

## RESULTS AND DISCUSSION

**Genuine topological Hall effect**

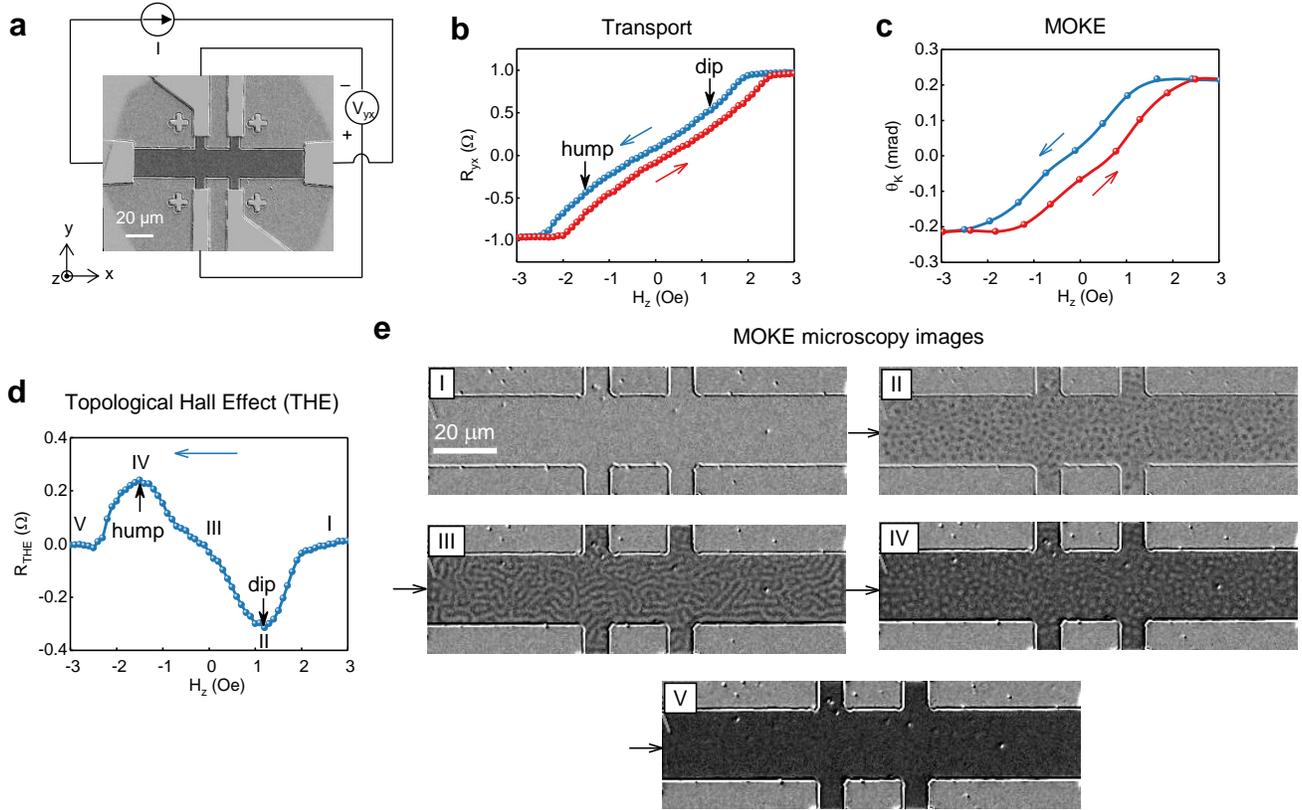

**Figure 1.** Genuine topological Hall effect (THE) + anomalous Hall effect (AHE). (a) An optical microscopy image of the device as well as the transport measurement setup. (b) The Hall resistance $R_{yx}$ as a function of an applied out-of-plane magnetic field $H_z$ at room temperature. The blue and red arrows indicate the direction of the field sweep, and the black arrows indicate the hump and dip caused by THE. (c) The Kerr rotation $\theta_K$ as a function of an applied out-of-plane magnetic field $H_z$ at room temperature. (d) The extracted THE component as a function of the applied magnetic field, with the same set of black arrows indicating the hump and dip as in (b). The Roman numerals indicate each phase of magnetic field sweep, and the blue and red arrows indicate the direction. (e)

Magneto-optical Kerr effect (MOKE) microscopy images corresponding to each phase of field sweep.

Here we are going to demonstrate the genuine THE induced by magnetic skyrmions. Both transport and magneto-optical Kerr effect (MOKE) measurements were performed at room temperature on the same Hall bar of Ta(5)/CoFeB(0.9)/Ir(0.03-0.15)/MgO(2)/Al$_2$O$_3$(5) as shown in Fig. 1a. A constant current $I$ was applied in the $x$ direction and the Hall voltage $V_{yx}$ was measured along $y$ direction, and the Hall resistance was obtained by $R_{yx} = V_{yx}/I$. An out-of-plane magnetic field $H_z$ was applied along $z$ direction. The Hall data as a function of $H_z$ as presented in Fig. 1b clearly show a magnetic hysteresis caused by the AHE. Meanwhile, the Kerr rotation data $\theta_K$ as a function of $H_z$ obtained by a laser MOKE in Fig. 1c also show a magnetic hysteresis as an indication of the ferromagnetic order, where the lines serving as a guide to the eye are cubic B-spline interpolations of the original data points. Strikingly, the slope of the transition region in the Hall data first decreases and then increases with the sweep of magnetic field, thus creating a hump and a dip as indicated by the black arrows. In sharp contrast, the Kerr rotation data are almost linear within the transition region and the slope does not change as much as the Hall data. The total Hall resistance is given by

$$R_{yx}(H_z) = R_0 H_z + R_S M_z(H_z) + R_{THE}(H_z) \qquad (3)$$

, where the first, second and third terms are ordinary, anomalous and topological Hall effect resistance. Since in a homogeneous ferromagnet, both the Kerr rotation and the AHE resistance are approximately linearly proportional to the magnetization normal to the plane, $M_z(H_z)$, or $R_{AHE}(H_z) \propto \theta_K(H_z) \propto M_z(H_z)$, the difference in the transition region is due to the additional contribution to the Hall resistance, the THE ($R_{THE}$). By subtracting the Kerr rotation data (scaled by a constant $\max[R_{yx}(H_z)] / \max[\theta_K(H_z)]$) from the Hall data, as well as the linear slope given by the ordinary Hall effect, we extract the topological Hall resistance as presented in Fig. 1d, exhibiting the same set of humps and dips as in Fig. 1b.

To confirm the origin of the THE, polar MOKE imaging experiments were carried out using a polar MOKE microscope to directly probe the magnetic domain structures, and the results are presented in Fig. 1e. With the sweep of an external magnetic field $H_z$, the magnetic domain structures evolve from a field-polarized uniform domain composed of up states only in phase I, to a skyrmion lattice with each skyrmion's center pointing down in phase II, to stripe domains where the up and down domains are almost equally distributed in phase III, to a skyrmion lattice with each skyrmion's center pointing up in phase IV, and finally to a field-polarized uniform domain composed of down states only in phase V. Notably, each phase corresponds to the topological Hall resistance well. Each skyrmion with its center pointing down and edge pointing up carries a topological charge of $Q_{sk} = -1$ regardless of being Néel-type or Bloch-type, [9,62] so the skyrmion lattice consisting of such skyrmions contributes to the negative dip of the topological Hall resistance in phase II. Similarly, each skyrmion with its center pointing down and edge pointing up carries a topological charge of $Q_{sk} = 1$, [9,62] and such skyrmion lattice contributes to the positive hump of the topological Hall resistance in phase IV. A uniform domain does not contribute to the topological Hall resistance in either phase I or phase V. In phase III, each stripe domain is topologically equivalent to and can be viewed as an elongated skyrmion. Since most of the stripes are connected and thus correspond to only a small number of skyrmions, they only give rise to a negligible THE.

In short, we have verified the coexistence of AHE and genuine THE induced by magnetic skyrmions through both transport and MOKE imaging, and an important feature of such signal is that the genuine THE occurs within the transition region of the AHE loop. In our material stack of Ta(5)/CoFeB(0.9)/Ir(0.03-0.15)/MgO(2)/$Al_2O_3$(5), the origin of the skyrmions is the interfacial Dzyaloshinskii-Moriya interactions (DMI) at the Ta/CoFeB interface, which is the same as our previous reports,[20,22,23] while the function of Ir wedge is to adjust the perpendicular magnetic anisotropy (PMA) of the ferromagnetic CoFeB layer. Our observations of genuine THE with AHE are also consistent with previous reports where magnetic skyrmions were directly observed by X-ray microscopy and magnetic force microscopy in Ir/Fe/Co/Pt multilayers. [31–33]

**Artifact "topological Hall effect"**

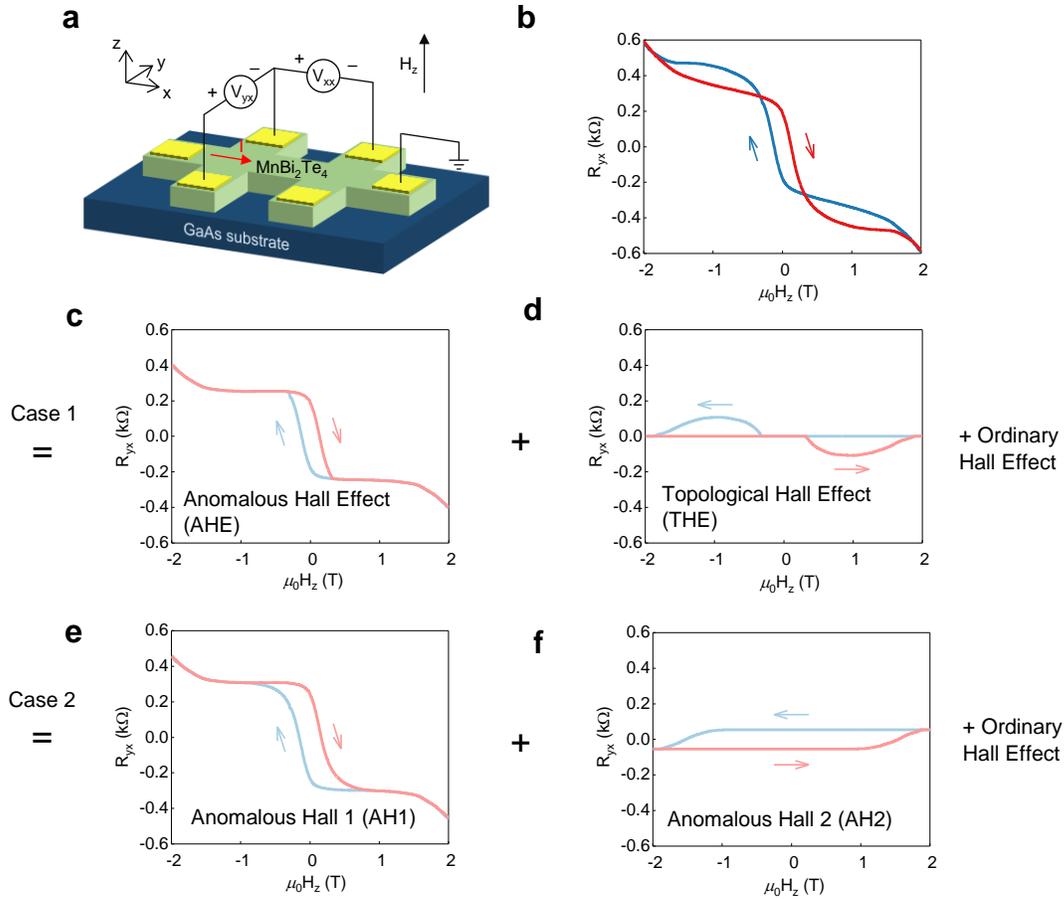

**Figure 2.** Artifact "THE" with AHE, or two-component AHE. (a) A schematic of the Hall bar device structure as well as the transport measurement setup. (b) The Hall resistance $R_{yx}$ as a function of an applied out-of-plane magnetic field $H_z$ at T=2 K. The blue and red arrows indicate the direction of the field sweep. (c), (d) For Case 1, the decomposition into anomalous Hall effect (AHE) + topological Hall effect (THE) + ordinary Hall effect (OHE). (e), (f) For Case 2, the decomposition into two AHEs (AH1 and AH2) + ordinary Hall effect (OHE).

Now we are going to demonstrate the artifact "THE", or two-component AHE. The transport measurements were performed at T=2 K on the 7 SL MnBi$_2$Te$_4$ Hall bar with a similar setup shown in Fig. 2a. The Hall data as a function of $H_z$ is presented in Fig. 2b, showing an AHE loop around zero field with non-monotonic features, or a hump and a dip that are similar to the THE outside the loop. In principle, such kind of Hall signal can always be hypothetically decomposed into either the sum of AHE and THE (Case 1 as illustrated in Fig. 2c and d) or two AHE loops, AH1 and AH2 (Case 2 as in Fig. 2e and f), plus a linear contribution from the ordinary Hall effect (OHE). If both AHE and THE were present as in the hypothesis of Case 1, the loop around zero field could be taken as the AHE, while the hump and dip outside the loop could be taken as the THE, which would be the most straightforward way of decomposition, and is actually used by a major part of previous literature reporting THE. [38,39,48–57,40–47] In the hypothesis of Case 2, two AHE loops were obtained from the decomposition. The hump and dip outside the loop could be taken as part of the positive AH2, and the rest of AH2 could be obtained by subtracting from the loop around zero field, which would yield the negative AH1. The detailed way of decomposition into two AHE loops can be found in the Section 3 of the supporting information. To express the two decompositions in equations,

$$\text{Case 1:} \quad R_{yx} = R_{AHE} + R_{THE} + R_0 H \qquad (4)$$

$$\text{Case 2:} \quad R_{yx} = R_{AH1} + R_{AH2} + R_0 H \qquad (5)$$

where $R_{AHE}$ and $R_{THE}$ are the Hall resistances of AHE and THE, $R_{AH1}$ and $R_{AH2}$ are the Hall resistances of AH1 and AH2, and the constant $R_0$ is the slope of the OHE.

At this point, both kinds of hypothetical decomposition seem reasonable, so that at first glance, the two-component AHE is hardly distinguishable from the genuine THE + AHE.

This has been exactly the problem that is troubling the transport studies of magnetic skyrmions with the THE alone as the evidence. [38,39,48–57,40–47] However, we will present three methods using: 1. Minor loops, 2. Temperature dependence, and 3. Gate dependence, to disprove the hypothesis of Case 1 (AHE+THE) and prove that this is actually two-component AHE as in Case 2, and further demonstrate that AH1 comes from $MnBi_2Te_4$ itself and the additional AHE component AH2 comes from the secondary phase $MnTe_2$ by both material characterizations and DFT calculations in the Section 1 & 2 of supporting information.

By comparing the artifact "THE" with AHE in Fig. 2 with the genuine one in Fig. 1, we could notice some important differences. In the case of artifact, the "THE" peaks at around 1.0 T and still persists to 1.8 T, while the "AHE component" is already fully saturated beyond the point of 0.3 T. Since the AHE is generally linearly proportional to the magnetization normal to the plane, chiral spin textures such as skyrmions would vanish and the magnetic structures would become a field-polarized uniform domain with such a large external magnetic field that is enough to fully saturate the AHE. In contrast, the genuine THE occurs in the transition region of the AHE, since MOKE microscopy reveals that chiral spin textures occur during the transition of magnetic domains from up to down states. To summarize, in the presence of AHE, it is more consistent for the THE to emerge within the transition region of the AHE loop rather than the saturation region.

**Method 1: Minor Loops**

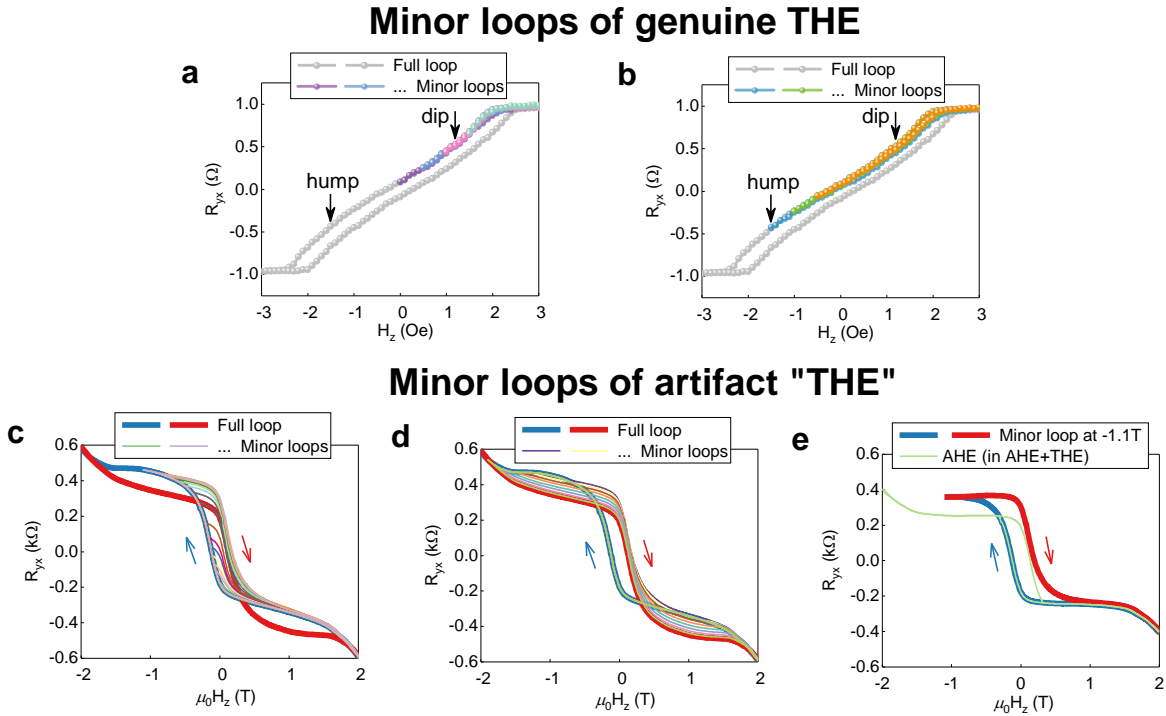

**Figure 3.** Minor loops. (a, b) Minor loops of genuine THE + AHE, with (a) showing minor loops stopping at +1.5 Oe, +1.0 Oe, +0.5 Oe and 0.0 Oe and (b) showing minor loops stopping at –0.5 Oe, –1.0 Oe and –1.5 Oe. (c-e) Minor loops of artifact "THE" with AHE, with (c) showing minor loops within the field of –1.1 T, (d) showing minor loops beyond the field of –1.2 T, and (e) showing the minor loop stopping at the field of –1.1 T (with OHE removed) and the "AHE component" under the assumption of AHE+THE.

To further explore the differences between genuine THE and the artifact, two-component AHE, our first method is minor loops. For genuine THE with AHE, the minor loops were taken with the same setup as in Fig. 1 by sweeping the magnetic field from +3.0 Oe to a certain value of magnetic field (+1.5 Oe, +1.0 Oe, +0.5 Oe … and –1.5 Oe) and then back to +3.0 Oe. As presented in Fig. 3a, for the minor loops stopping at +1.5 Oe, +1.0 Oe, +0.5 Oe and 0.0 Oe, the forward and backward sweeping curves basically collapse together and follow the forward sweeping curve of the full loop due to the weak perpendicular magnetic anisotropy (PMA) of the sample, while the minor loops stopping at –0.5 Oe, –1.0 Oe and –1.5 Oe in Fig. 3b sequentially develops hysteresis within the full loop. Notably, all the minor loops are within the full loop and do not reveal a minor anomalous Hall loop that exceeds the full loop. Moreover, they largely follow the trajectory of the full loop and exhibit the same set of humps and dips that do not depend on the history of sweeping magnetic field. This is consistent with the observations of chiral spin textures.

For artifact "THE" with AHE, the minor loops were taken on the same Hall bar as in Fig. 2 at T=2 K by sweeping the magnetic field from +2 T to zero or a certain value of negative field (0 T, –0.05 T, –0.075 T, –0.1 T, –0.125 T, –0.15 T, –0.2 T, –0.3 T, –0.4 T, ... –1.9 T, –2.0 T) and then back to +2 T. Minor loops within the field of –1.1 T, as presented in Fig. 3c, do not reach the field range of the humps at negative fields and show no dip at positive fields, while those beyond the field of –1.2 T, as presented in Fig. 3d, gradually develop dips at positive fields as well as humps at negative fields simultaneously. The very fact that emergence of humps and dips depend on the history of sweeping magnetic field reveals that the humps and dips originate from the sum of two AHE loops, because within the field of –1.1 T, the AH2, which has a positive polarity and contributes to the humps and dips in the negative AH1, remains unswitched and thus does not contribute to the humps and dips at all. Meanwhile, beyond the field of –1.2 T, the AH2 starts being partially and finally fully switched, thus gradually giving rise to the humps and dips.

More importantly, the minor loop that stopped at –1.1 T (with OHE removed), as highlighted in Fig. 3e, clearly cannot fit into the full "AHE component" under the assumption that the Hall signal were AHE+THE. This contradiction is the decisive evidence against the assumption of AHE+THE, thus supporting that such Hall signal is the sum of two AHE loops.

In conclusion, the minor loops of the genuine THE with AHE always reside within the full loop, while those of the artifact "THE" may reveal a single loop that cannot fit into the "AHE component". Besides, for the genuine THE with AHE, the emergence of humps and dips in the minor loops do not depend on the history of sweeping magnetic field, while the opposite applies for the artifact.

**Method 2: Temperature dependence**

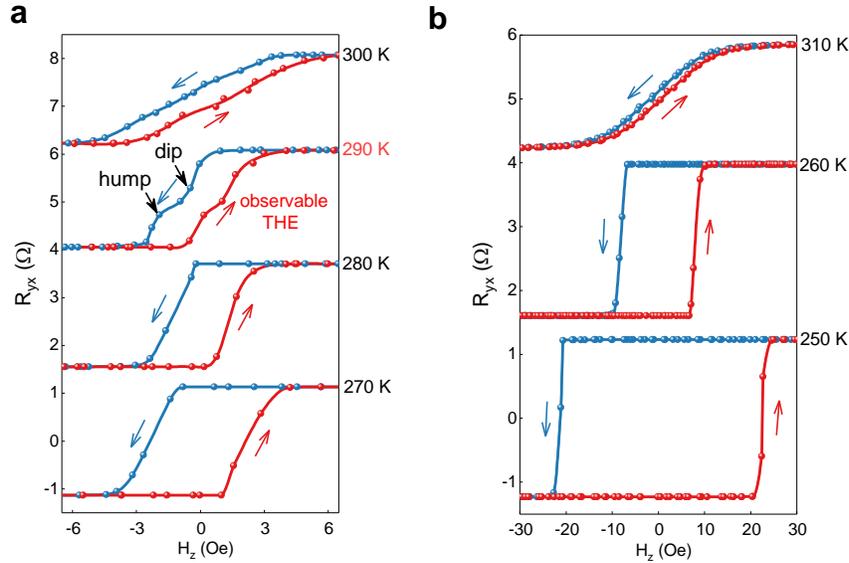

**Figure 4.** The temperature dependence of genuine THE with AHE. The Hall resistance $R_{yx}$ as a function of an applied out-of-plane magnetic field $H_z$ under (a) 270 K, 280 K, 290 K and 300 K, with a zoomed-in field range of $\pm 6.5$ Oe and (b) 250 K, 260 K and 310 K, with a greater field range of $\pm 30$ Oe. The data at 280 K, 290 K, 300 K, 260 K and 310 K are shifted vertically for clarity, and the blue and red arrows indicating the direction of the field sweep.

Our second method of distinguishing genuine THE from the artifact, two-component AHE is temperature dependence. The temperature dependence of genuine THE with AHE was investigated in the same device as in Fig. 1. While the transport data presented in Fig. 1 was taken at room temperature using a normal Helmholtz coil with a field resolution of 0.1 Oe, the temperature dependence data were taken using a 9 T superconducting magnet, of which the finest resolution is only ~ 1 Oe. The Hall resistance as a function of an applied out-of-plane magnetic field $H_z$ taken under various temperatures is presented in Fig. 4. Here the original data points are presented with lines serving as a guide to the eye, which are either cubic B-spline (with smoothing) or Akima spline interpolation of the original data points.

From the temperature dependence shown in Fig. 4a and b, we note that only the Hall data at 290 K have non-monotonic features of humps and dips that are observable and indicate the THE, since the perpendicular magnetic anisotropy (PMA) of the sample changes dramatically with temperature. At 310 K, the anomalous Hall loop is S-shaped and requires an external field as large as 20 Oe to saturate the Hall resistance and thus align all the spins, indicating weak magnetic anisotropy. At 300 K, the anomalous Hall loop is still S-shaped but the external field required to saturate the Hall resistance shrinks to 6 Oe. Notably for both 310 K and 300 K, the Hall resistance is quite monotonic (or even linear) with regard to magnetic field within the transition region, and no hump or dip can be resolved. As temperature decreases and magnetic anisotropy gets stronger, strikingly at 290 K, non-monotonic features of humps and dips from the THE emerge within the transition region of the AHE, which is consistent with the room-temperature Hall data in Fig. 1b (here the separation between forward and backward sweeping curves might be exaggerated due to superconducting magnet remanence). When temperature further decreases below 280 K, the AHE loops becomes more square-like as the transition becomes increasingly sharper, and the coercive field keeps increasing as well, indicating stronger magnetic anisotropy. With the field resolution of 1 Oe, the number of available data points within the transition region is not enough to prove the existence of THE. Since the energy terms responsible for rendering possible the existence of magnetic skyrmions include the exchange interaction,

Dzyaloshinskii-Moriya interaction (DMI), Zeeman energy and magnetic anisotropy energy, [9,10] it is understandable that the highly changeable PMA of CoFeB with regard to temperature limits the emergence of the THE to a small temperature range, but this is largely sample-dependent and not generalizable, as it is possible for magnetic skyrmions to keep increasing density when lowering temperature. [9,10] We will discuss the general features for distinguishing genuine THE from the artifact by comparing with the temperature dependence of artifact "THE" or two-component AHE in the latter half of this section.

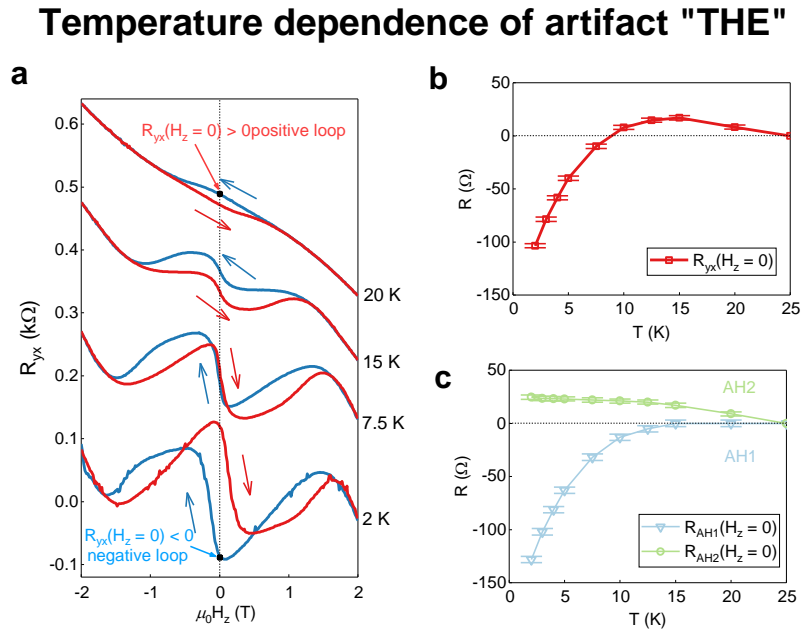

**Figure 5.** The temperature dependence of artifact "THE", or two-component AHE. (a) The Hall resistance $R_{yx}$ as a function of an applied out-of-plane magnetic field $H_z$ under various temperatures, with blue and red arrows indicating the direction of the field sweep. The data at T= 7.5 K, 15 K and 20 K are shifted vertically for clarity. (b) The temperature

dependence of the zero-field Hall resistance $R_{yx}(H_z = 0)$ taken after sweeping the magnetic field from +2 T to 0 T (c) The temperature dependence of two fitted AHE components at zero field, $R_{AH1}(H_z = 0)$ and $R_{AH2}(H_z = 0)$.

Here we present the temperature dependence of artifact "THE", or two-component AHE, which was measured in a Hall bar of 5 SL MnBi$_2$Te$_4$ sample. The Hall resistance as a function of an applied out-of-plane magnetic field $H_z$ was taken under various temperatures, as presented in Fig. 5a. In the Fig. 5b, the zero-field Hall resistance $R_{yx}(H_z = 0)$ taken after sweeping the magnetic field from +2 T to 0 T is extracted as a function of temperature, and the polarity of the Hall loop is defined as the sign of $R_{yx}(H_z = 0)$ (positive or negative). The $R_{yx}(H_z = 0)$ turns from negative to positive with increasing temperature, thus indicating a polarity change of the Hall loop from negative to positive, and eventually vanishes above 25 K, which is consistent with the previous reports of the Néel temperature of MnBi$_2$Te$_4$.[63–67]

Strikingly, the Hall signal evolves from a negative loop with a hump and a dip to a single positive loop by increasing the temperature. As mentioned before, if the presence of AHE and THE were assumed, the loop around zero field could be taken as the "AHE component", while the hump and the dip outside the loop could be taken as the "THE component". As the temperature increases, the hump and the dip that could be taken as the "THE component" start to get closer to the zero field at 7.5 K, merge together and flip the polarity of the Hall signal from negative to positive with some non-monotonic features preserved at 15 K, and eventually form a single positive loop and lose all their non-monotonic characteristics at 20 K (when the linear background is removed). Such a continuous evolution of the hump and the dip with temperature reveals their origin is the positive AHE loop, or AH2 in the decomposition into two AHEs.

The fitted results of the two AHE components at zero field are presented in Fig. 5c, and the full loops are presented in the Fig. S7 of the supporting information. With the increase of temperature, $R_{AH1}(H_z = 0)$ decreases quickly and becomes zero beyond 15 K, while $R_{AH2}(H_z = 0)$ follows a much slower trend of decrease and eventually vanishes above 25 K. The difference in the temperature dependences of AH1 and AH2, which carry opposite signs, explains the polarity change of the Hall signal with temperature.

Therefore, such emergence of humps and dips with a polarity change in the Hall signal is an important feature of the artifact "THE", or two-component AHE. In comparison, such feature of polarity change with temperature is absent in the case of genuine THE with AHE shown in Fig. 4. This is because genuine THE only occurs within the transition region of the AHE and should not affect the polarity of the AHE loop.

**Method 3: Gate dependence**

## Gate dependence of artifact "THE"

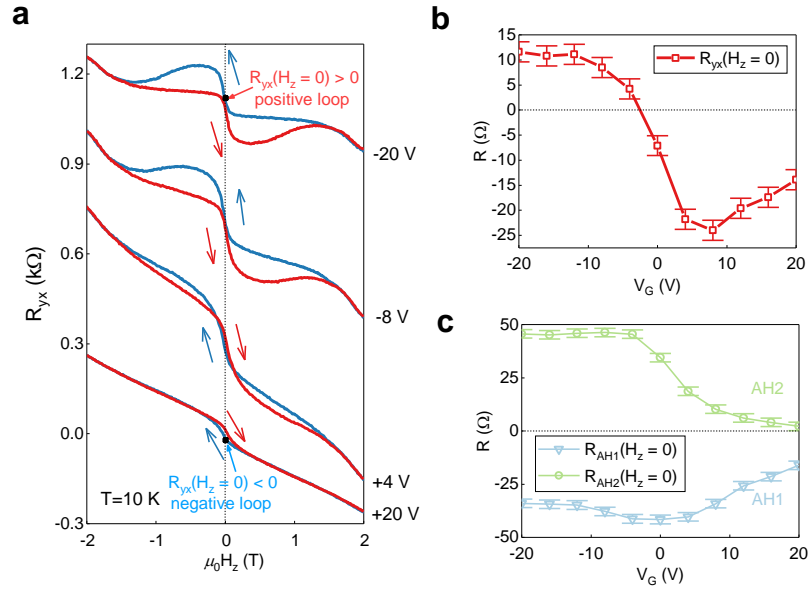

**Figure 6.** The gate dependence of artifact "THE", or two-component AHE. (a) The Hall resistance $R_{yx}$ as a function of an applied out-of-plane magnetic field $H_z$ under various gate biases, with blue and red arrows indicating the direction of the field sweep. The data at $V_G$= +4 V, –8 V and –20 V are shifted vertically for clarity. (b) The gate dependence of the zero-field Hall resistance $R_{yx}(H_z = 0)$ (c) The gate dependence of two fitted AHE components at zero field, $R_{AH1}(H_z = 0)$ and $R_{AH2}(H_z = 0)$.

Our third method of distinguishing genuine THE from the artifact, two-component AHE is gate dependence. The gate dependence of artifact "THE", or two-component AHE was obtained in another Hall bar of the 7 SL MnBi$_2$Te$_4$ sample with a top gate at T=10 K, as presented in Fig. 6a. In Fig. 6b, the zero-field Hall resistance $R_{yx}(H_z = 0)$ taken after sweeping the magnetic field from +2 T to 0 T is extracted as a function of gate voltage. The $R_{yx}(H_z = 0)$ turns from positive to negative with increasing gate bias, indicating a polarity change of the Hall loop from positive to negative. The Hall conductivity $\sigma_{xy}(H_z =$

0) carries the same sign as $R_{yx}(H_z = 0)$, and its gate dependence is also presented in Fig. S9 of the supporting information to compare with the DFT calculations in Fig. S3 of the supporting information.

By increasing the gate bias, the Hall signal evolves from a positive loop with a hump and a dip to a single negative loop, which is similar to the temperature dependence discussed previously. Therefore, a similar line of reasoning also applies to the gate dependence. As the gate bias is increased from –20 V to –8 V, the hump and the dip, or the "THE component" start to shrink and the Hall signal keeps the same polarity. At +4 V, as the hump and the dip further shrink, the Hall signal flips its polarity from positive to negative dramatically. When the gate bias is further increased to +20 V, the hump and the dip almost vanish and only a single negative AHE loop is left, which is the AH1 from the decomposition into two AHEs. This also corroborates the existence of a negative AH1 component, which gives rise to the non-monotonic characteristics of the Hall signal around zero field. When AH1 is removed from the Hall signals, only a positive AH2 is left, and the non-monotonic features are gone. (see Fig. S8 of the supporting information for detailed decompositions)

The fitted results of the two AHE components at zero field are plotted against gate bias in Fig. 6c, and the full loops are presented in the Fig. S8 of the supporting information. With the increase of gate bias, $R_{AH1}(H_z = 0)$ follows an ambipolar response to the gate voltage and changes by a factor of 4.1 in amplitude only, while $R_{AH2}(H_z = 0)$ almost saturates at negative gate voltage and decreases rapidly at positive gate voltage by a factor of 16.7. The obvious difference in the gate dependence of AH1 and AH2, which carry opposite signs, also accounts for the polarity change of the Hall signal with gate.

Therefore, similar to our method 2 of temperature dependence, we also show from the gate dependence that such emergence of humps and dips with a polarity change in the Hall signal distinguishes artifact "THE", or two-component AHE from genuine THE.

CONCLUSIONS

In conclusion, we have confirmed genuine THE with AHE by both magnetotransport and MOKE microscopy, and magnetic skyrmions have been directly observed. By comparing the genuine THE with AHE with the artifact, or two-component AHE, we have found that genuine THE is present within the transition region of the AHE loop, while the artifact "THE", or two-component AHE can even persist well beyond the saturation of the "AHE component" under the false assumption of AHE+THE. Furthermore, we have proposed three methods: 1. Minor loops, 2. Temperature dependence, 3. Gate dependence for distinguishing the artifact from genuine THE. By carefully examining the minor loops, we found that for artifact "THE", or two-component AHE, one of the minor loops is a single loop that cannot fit into the "AHE component", while minor loops of genuine THE with AHE are always within the full loop. In addition, by either tuning the temperature or the gate bias in artifact "THE", or two-component AHE, the non-monotonic features, or the humps and dips could vanish and simultaneously the AHE loop undergoes a polarity change. Meanwhile, the temperature dependence of genuine THE with AHE reveals no such polarity change of the AHE, since genuine THE in the transition region does not affect the polarity of the AHE loop. Our work shows that caution must be exercised in determining whether the non-monotonic features, or humps and dips in the Hall signals are truly THE or not.

METHODS

**Material growth and device fabrication.** Layers consisting of Ta(5)/CoFeB(0.9)/Ir(0.03-0.15)/MgO(2)/Al$_2$O$_3$(5) were grown on Si/SiO$_2$ substrates by D.C. and radiofrequency magnetron sputtering at room temperature (numbers in parentheses represent thickness in nm). The Ir layer has a wedge shape with continuously changing thickness with a nominal gradient of 0.02 nm per 1 cm of the sample length. The films were then annealed at 250 °C for 30 minutes to enhance their PMA. The samples were then patterned into Hall bar devices using standard photolithography techniques. The size of the Hall bars was 20 μm × 130 μm. Cr/Au contact electrodes with thicknesses of 10/100 nm were deposited using an electron beam evaporator.

The MnBi$_2$Te$_4$ samples were grown in an ultra-high vacuum, Perkin-Elmer molecular beam epitaxy (MBE) system. We used epi-ready semi-insulating GaAs (111)B substrates for the growth. The entire process was monitored by the reflection high-energy electron diffraction (RHEED) *in situ*, where the digital RHEED images were captured using a KSA400 system built by K-space Associates, Inc. Before growth, the substrates were loaded into the MBE chamber and pre-annealed at the temperature of 630 °C in a Te-rich environment to desorb the oxide on the surface. During growth, we kept the substrate at 200 °C. High-purity Mn, Bi and Te were evaporated simultaneously from standard Knudsen cells. After the deposition, the film was post-annealed in a Te-rich environment at 290 °C for 2 minutes to improve crystallinity. The sharp and streaky lines in the RHEED pattern indicate good epitaxial crystal quality. Two types of Hall bar devices were fabricated to characterize the transport properties. The first type without a gate was millimeter-sized and made for the samples in Fig. 2, 3 & 5. The thin films were fabricated into Hall bar patterns with the dimensions of 2 mm (length) × 1 mm (width) using standard etching with a hard mask. Indium contacts were later soldered onto the devices. The second type with a top gate was micron-sized and made for the sample in Fig. 6. The thin film was fabricated into a 40 μm (length) × 40 μm (width) Hall bar geometry using a standard photolithography process. Cr/Au contact electrodes with thicknesses of 10/100 nm were deposited using an electron beam evaporator. Mica and graphite thin flakes, which serve as the gate dielectric and the electrode, respectively, were exfoliated and subsequently transferred onto the as-fabricated Hall bar device for top gating.

**Transport measurements.** Two types of transport measurements were carried out. The room-temperature magnetoelectric transport measurements presented in Fig. 1 & 3 were conducted with a Karl Suss PA200 Semiautomatic probe station, which is equipped with a Helmholtz coil driven by a Kepco power supply to generate an external magnetic field. A Keithley 2636 source-meter was used for supplying a DC source current as well as measuring the Hall voltage.

Low temperature magnetoelectric transport measurements presented in Fig. 2 - 6 were conducted in a Quantum Design physical property measurement system (PPMS) equipped with a 9 T superconducting magnet with a base temperature of 1.9 K. A Keithley 6221 current source was used to generate a source AC current, and multiple lock-in amplifiers (Stanford Research SR830) were used to obtain the Hall voltage from the Hall bar devices. Gate voltage was applied to the gate electrode of the $MnBi_2Te_4$ Hall bar device with a top gate in Fig. 6 by a Keithley 2636 source meter.

**Magneto-optical Kerr Effect (MOKE) measurements.** Two types of MOKE measurements were carried out. The Kerr rotation data in Fig. 1(c) was obtained with a 633 nm He-Ne continuous-wave laser modulated by a photoelastic modulator (PEM). The setup was based on the polar MOKE configuration where the fabricated device was perpendicular to the optical beam. The reflected light was separated into p- and s- waves by a Wollaston prism and collected by a balanced photodiode. The signal was then transmitted to a lock-in amplifier (Stanford Research SR830) and stored in the computer. The external magnetic field was generated by a Helmholtz coil driven by a Kepco power supply.

For the MOKE microscopy images in Fig. 1(e), the polar MOKE imaging experiments were carried out using a spatially (360 nm resolution) and temporally (20 ms resolution) resolved polar MOKE microscope. The external magnetic field was also generated by a Helmholtz coil driven by a Kepco power supply.

AUTHOR INFORMATION

**Corresponding Author**

* Correspondence to: wang@ee.ucla.edu, wur@ucr.edu

**Author Contributions**

† L.T., B.D. and J.L. contributed equally to this work.

L.T. and K.L.W. conceived and designed the experiments. K.L.W. supervised the work. For the Ta(5)/CoFeB(0.9)/Ir(0.03-0.15)/MgO(2)/Al$_2$O$_3$(5) sample with skyrmions, B.D. grew the sample, K.W. fabricated it into Hall bar devices, L.T. and B.D. carried out the transport measurements, H.H. performed the laser MOKE measurement, and B.D. performed the MOKE microscopy measurement. For the MnBi$_2$Te$_4$ samples, L.T. grew the samples, performed the X-ray diffraction, and fabricated the millimeter sized Hall bar devices, S.K.C. fabricated the micron sized Hall bar device and transferred the top gate, L.T. and S.K.C. carried out the transport experiments, and H.Z. and A.V.D. performed the transmission electron microscopy measurement. L.T. processed all the data. J. L. and R.W. performed the DFT calculations. All authors contributed to the analyses. L.T., B.D., J.L. and K.L.W. wrote the manuscript with contributions from all authors.

**Notes**

A previous version of this manuscript has been submitted to a preprint server.[68] The authors declare no competing financial interest.

ACKNOWLEDGMENT

The authors acknowledge the support from the National Science Foundation (NSF) (DMR-1411085 and DMR-1810163) and the Army Research Office Multidisciplinary University Research Initiative (MURI) under grant numbers W911NF16-1-0472 and W911NF-19-S-0008. In addition, H.Z. acknowledges support from the U.S. Department of Commerce, NIST under financial assistance award 70NANB19H138. A.V.D. acknowledges support from the Material Genome Initiative funding allocated to NIST. DFT calculations by J. L. and R. W. were supported by US DOE, Basic Energy Science (Grant No. DE-FG02-05ER46237) and NERSC. C.E. is an employee of Fibertek, Inc. and performs in support of Contract No.W15P7T19D0038, Delivery Order W911-QX-20-F-0023. The views expressed are those of the authors and do not reflect the official policy or position of the Department of Defense or the US government. The identification of any

ASSOCIATED CONTENT

**Supporting Information Available:** 1. Material characterizations of $MnBi_2Te_4$, 2. The origins of two AHE components by DFT calculations, 3. The method of AHE decompositions, 4. Gate dependence of Hall conductivity, 5. Thickness dependence, 6. Reproducibility and controllability.

REFERENCES

(1) Nagaosa, N.; Sinova, J.; Onoda, S.; MacDonald, A. H.; Ong, N. P. Anomalous Hall Effect. *Rev. Mod. Phys.* **2010**, *82* (2), 1539–1592. https://doi.org/10.1103/RevModPhys.82.1539.

(2) Bruno, P.; Dugaev, V. K.; Taillefumier, M. Topological Hall Effect and Berry Phase in Magnetic Nanostructures. *Phys. Rev. Lett.* **2004**, *93* (9), 096806. https://doi.org/10.1103/PhysRevLett.93.096806.

(3) Smit, J. The Spontenous Hall Effect in Ferromagnetics I. *Physica* **1955**, *21*, 877–887. https://doi.org/10.1016/S0031-8914(55)92596-9.

(4) Smit, J. The Spontenous Hall Effect in Ferromagnetics II. *Physica* **1958**, *24* (1–5), 39–51. https://doi.org/10.1016/S0031-8914(58)93541-9.

(5) Berger, L. Side-Jump Mechanism for the Hall Effect of Ferromagnets. *Phys. Rev. B* **1970**, *2* (11), 4559–4566. https://doi.org/10.1103/PhysRevB.2.4559.

(6)　Berger, L. Application of the Side-Jump Model to the Hall Effect and Nernst Effect in Ferromagnets. *Phys. Rev. B* **1972**, *5* (5), 1862–1870. https://doi.org/10.1103/PhysRevB.5.1862.

(7)　Jungwirth, T.; Niu, Q.; MacDonald, A. H. Anomalous Hall Effect in Ferromagnetic Semiconductors. *Phys. Rev. Lett.* **2002**, *88* (20), 207208. https://doi.org/10.1103/PhysRevLett.88.207208.

(8)　Ye, J.; Kim, Y. B.; Millis, A. J.; Shraiman, B. I.; Majumdar, P.; Teanovic, Z. Berry Phase Theory of the Anomalous Hall Effect: Application to Colossal Magnetoresistance Manganites. *Phys. Rev. Lett.* **1999**, *83* (18), 3737–3740. https://doi.org/10.1103/PhysRevLett.83.3737.

(9)　Nagaosa, N.; Tokura, Y. Topological Properties and Dynamics of Magnetic Skyrmions. *Nat. Nanotechnol.* **2013**, *8* (12), 899–911. https://doi.org/10.1038/nnano.2013.243.

(10)　Fert, A.; Reyren, N.; Cros, V. Magnetic Skyrmions: Advances in Physics and Potential Applications. *Nat. Rev. Mater.* **2017**, *2*, 17031. https://doi.org/10.1038/natrevmats.2017.31.

(11)　Dzyaloshinsky, I. A Thermodynamic Theory of "Weak" Ferromagnetism of Antiferromagnetics. *J. Phys. Chem. Solids* **1958**, *4* (4), 241–255. https://doi.org/10.1016/0022-3697(58)90076-3.

(12)　Moriya, T. Anisotropic Superexchange Interaction and Weak Ferromagnetism. *Phys. Rev.* **1960**, *120* (1), 91–98. https://doi.org/10.1103/PhysRev.120.91.

(13) Garel, T.; Doniach, S. Phase Transitions with Spontaneous Modulation-the Dipolar Ising Ferromagnet. *Phys. Rev. B* **1982**, *26* (1), 325–329. https://doi.org/10.1103/PhysRevB.26.325.

(14) Okubo, T.; Chung, S.; Kawamura, H. Multiple-q States and the Skyrmion Lattice of the Triangular-Lattice Heisenberg Antiferromagnet under Magnetic Fields. *Phys. Rev. Lett.* **2012**, *108*, 017206. https://doi.org/10.1103/PhysRevLett.108.017206.

(15) Rößler, U. K.; Bogdanov, A. N.; Pfleiderer, C. Spontaneous Skyrmion Ground States in Magnetic Metals. *Nature* **2006**, *442* (7104), 797–801. https://doi.org/10.1038/nature05056.

(16) Mühlbauer, S.; Binz, B.; Jonietz, F.; Pfleiderer, C.; Rosch, A.; Neubauer, A.; Georgii, R.; Böni, P. Skyrmion Lattice in a Chiral Magnet. *Science* **2009**, *323* (5916), 915-919. https://doi.org/10.1126/science.1166767.

(17) Yu, X. Z.; Onose, Y.; Kanazawa, N.; Park, J. H.; Han, J. H.; Matsui, Y.; Nagaosa, N.; Tokura, Y. Real-Space Observation of a Two-Dimensional Skyrmion Crystal. *Nature* **2010**, *465* (7300), 901–904. https://doi.org/10.1038/nature09124.

(18) Seki, S.; Yu, X. Z.; Ishiwata, S.; Tokura, Y. Observation of Skyrmions in a Multiferroic Material. *Science* **2012**, *336* (6078), 198–201. https://doi.org/10.1126/science.1214143.

(19) Milde, P.; Köhler, D.; Seidel, J.; Eng, L. M.; Bauer, A.; Chacon, A.; Kindervater, J.; Mühlbauer, S.; Pfleiderer, C.; Buhrandt, S.; Schütte, C.; Rosch, A. Unwinding of a Skyrmion Lattice by Magnetic Monopoles. *Science* **2013**, *340* (6136), 1076–1080. https://doi.org/10.1126/science.1234657.

(20) Jiang, W.; Upadhyaya, P.; Zhang, W.; Yu, G.; Jungfleisch, M. B.; Fradin, F. Y.; Pearson, J. E.; Tserkovnyak, Y.; Wang, K. L.; Heinonen, O.; Te Velthuis, S. G. E.; Hoffmann, A. Blowing Magnetic Skyrmion Bubbles. *Science* **2015**, *349* (6245), 283–286. https://doi.org/10.1126/science.aaa1442.

(21) Miao, B. F.; Sun, L.; Wu, Y. W.; Tao, X. D.; Xiong, X.; Wen, Y.; Cao, R. X.; Wang, P.; Wu, D.; Zhan, Q. F.; You, B.; Du, J.; Li, R. W.; Ding, H. F. Experimental Realization of Two-Dimensional Artificial Skyrmion Crystals at Room Temperature. *Phys. Rev. B* **2014**, *90*, 174411. https://doi.org/10.1103/PhysRevB.90.174411.

(22) Jiang, W.; Zhang, X.; Yu, G.; Zhang, W.; Wang, X.; Benjamin Jungfleisch, M.; Pearson, J. E.; Cheng, X.; Heinonen, O.; Wang, K. L.; Zhou, Y.; Hoffmann, A.; Te Velthuis, S. G. E. Direct Observation of the Skyrmion Hall Effect. *Nat. Phys.* **2017**, *13* (2), 162–169. https://doi.org/10.1038/nphys3883.

(23) Yu, G.; Upadhyaya, P.; Shao, Q.; Wu, H.; Yin, G.; Li, X.; He, C.; Jiang, W.; Han, X.; Amiri, P. K.; Wang, K. L. Room-Temperature Skyrmion Shift Device for Memory Application. *Nano Lett.* **2017**, *17* (1), 261–268. https://doi.org/10.1021/acs.nanolett.6b04010.

(24) Langner, M. C.; Roy, S.; Mishra, S. K.; Lee, J. C. T.; Shi, X. W.; Hossain, M. A.; Chuang, Y. D.; Seki, S.; Tokura, Y.; Kevan, S. D.; Schoenlein, R. W. Coupled Skyrmion Sublattices in Cu2OSeO3. *Phys. Rev. Lett.* **2014**, *112*, 167202.

https://doi.org/10.1103/PhysRevLett.112.167202.

(25) Li, J.; Tan, A.; Moon, K. W.; Doran, A.; Marcus, M. A.; Young, A. T.; Arenholz, E.; Ma, S.; Yang, R. F.; Hwang, C.; Qiu, Z. Q. Tailoring the Topology of an Artificial Magnetic Skyrmion. *Nat. Commun.* **2014**, *5*, 4704. https://doi.org/10.1038/ncomms5704.

(26) Neubauer, A.; Pfleiderer, C.; Binz, B.; Rosch, A.; Ritz, R.; Niklowitz, P. G.; Böni, P. Topological Hall Effect in the A Phase of MnSi. *Phys. Rev. Lett.* **2009**, *102*, 186602. https://doi.org/10.1103/PhysRevLett.102.186602.

(27) Lee, M.; Kang, W.; Onose, Y.; Tokura, Y.; Ong, N. P. Unusual Hall Effect Anomaly in MnSi under Pressure. *Phys. Rev. Lett.* **2009**, *102*, 186601. https://doi.org/10.1103/PhysRevLett.102.186601.

(28) Yu, X. Z.; Kanazawa, N.; Onose, Y.; Kimoto, K.; Zhang, W. Z.; Ishiwata, S.; Matsui, Y.; Tokura, Y. Near Room-Temperature Formation of a Skyrmion Crystal in Thin-Films of the Helimagnet FeGe. *Nat. Mater.* **2011**, *10* (2), 106–109. https://doi.org/10.1038/nmat2916.

(29) Huang, S. X.; Chien, C. L. Extended Skyrmion Phase in Epitaxial FeGe(111) Thin Films. *Phys. Rev. Lett.* **2012**, *108*, 267201. https://doi.org/10.1103/PhysRevLett.108.267201.

(30) Li, Y.; Kanazawa, N.; Yu, X. Z.; Tsukazaki, A.; Kawasaki, M.; Ichikawa, M.; Jin, X. F.; Kagawa, F.; Tokura, Y. Robust Formation of Skyrmions and Topological Hall Effect Anomaly in Epitaxial Thin Films of MnSi. *Phys. Rev. Lett.* **2013**, *110*,

117202. https://doi.org/10.1103/PhysRevLett.110.117202.

(31) Soumyanarayanan, A.; Raju, M.; Oyarce, A. L. G.; Tan, A. K. C.; Im, M. Y.; Petrovic, A. P.; Ho, P.; Khoo, K. H.; Tran, M.; Gan, C. K.; Ernult, F.; Panagopoulos, C. Tunable Room-Temperature Magnetic Skyrmions in Ir/Fe/Co/Pt Multilayers. *Nat. Mater.* **2017**, *16* (9), 898–904. https://doi.org/10.1038/NMAT4934.

(32) Raju, M.; Yagil, A.; Soumyanarayanan, A.; Tan, A. K. C.; Almoalem, A.; Ma, F.; Auslaender, O. M.; Panagopoulos, C. The Evolution of Skyrmions in Ir/Fe/Co/Pt Multilayers and Their Topological Hall Signature. *Nat. Commun.* **2019**, *10*, 696. https://doi.org/10.1038/s41467-018-08041-9.

(33) Raju, M.; Petrović, A. P.; Yagil, A.; Denisov, K. S.; Duong, N. K.; Göbel, B.; Şaşıoğlu, E.; Auslaender, O. M.; Mertig, I.; Rozhansky, I. V.; Panagopoulos, C. Colossal Topological Hall Effect at the Transition between Isolated and Lattice-Phase Interfacial Skyrmions. *Nat. Commun.* **2021**, *12*, 2758. https://doi.org/10.1038/s41467-021-22976-6.

(34) Kurumaji, T.; Nakajima, T.; Hirschberger, M.; Kikkawa, A.; Yamasaki, Y.; Sagayama, H.; Nakao, H.; Taguchi, Y.; Arima, T. hisa; Tokura, Y. Skyrmion Lattice with a Giant Topological Hall Effect in a Frustrated Triangular-Lattice Magnet. *Science* **2019**, *365* (6456), 914–918. https://doi.org/10.1126/science.aau0968.

(35) Vistoli, L.; Wang, W.; Sander, A.; Zhu, Q.; Casals, B.; Cichelero, R.; Barthélémy, A.; Fusil, S.; Herranz, G.; Valencia, S.; Abrudan, R.; Weschke, E.; Nakazawa, K.;

Kohno, H.; Santamaria, J.; Wu, W.; Garcia, V.; Bibes, M. Giant Topological Hall Effect in Correlated Oxide Thin Films. *Nat. Phys.* **2019**, *15* (1), 67–72. https://doi.org/10.1038/s41567-018-0307-5.

(36) Wu, Y.; Zhang, S.; Zhang, J.; Wang, W.; Zhu, Y. L.; Hu, J.; Yin, G.; Wong, K.; Fang, C.; Wan, C.; Han, X.; Shao, Q.; Taniguchi, T.; Watanabe, K.; Zang, J.; Mao, Z.; Zhang, X.; Wang, K. L. Néel-Type Skyrmion in WTe2/Fe3GeTe2 van Der Waals Heterostructure. *Nat. Commun.* **2020**, *11*, 3860. https://doi.org/10.1038/s41467-020-17566-x.

(37) Maccariello, D.; Legrand, W.; Reyren, N.; Garcia, K.; Bouzehouane, K.; Collin, S.; Cros, V.; Fert, A. Electrical Detection of Single Magnetic Skyrmions in Metallic Multilayers at Room Temperature. *Nat. Nanotechnol.* **2018**, *13* (3), 233–237. https://doi.org/10.1038/s41565-017-0044-4.

(38) Ishiwata, S.; Tokunaga, M.; Kaneko, Y.; Okuyama, D.; Tokunaga, Y.; Wakimoto, S.; Kakurai, K.; Arima, T.; Taguchi, Y.; Tokura, Y. Versatile Helimagnetic Phases under Magnetic Fields in Cubic Perovskite SrFeO3. *Phys. Rev. B* **2011**, *84*, 054427. https://doi.org/10.1103/PhysRevB.84.054427.

(39) Kanazawa, N.; Kubota, M.; Tsukazaki, A.; Kozuka, Y.; Takahashi, K. S.; Kawasaki, M.; Ichikawa, M.; Kagawa, F.; Tokura, Y. Discretized Topological Hall Effect Emerging from Skyrmions in Constricted Geometry. *Phys. Rev. B* **2015**, *91*, 041122(R). https://doi.org/10.1103/PhysRevB.91.041122.

(40) Ohuchi, Y.; Kozuka, Y.; Uchida, M.; Ueno, K.; Tsukazaki, A.; Kawasaki, M. Topological Hall Effect in Thin Films of the Heisenberg Ferromagnet EuO. *Phys.*

Rev. B **2015**, *91*, 245115. https://doi.org/10.1103/PhysRevB.91.245115.

(41) Yasuda, K.; Wakatsuki, R.; Morimoto, T.; Yoshimi, R.; Tsukazaki, A.; Takahashi, K. S.; Ezawa, M.; Kawasaki, M.; Nagaosa, N.; Tokura, Y. Geometric Hall Effects in Topological Insulator Heterostructures. *Nat. Phys.* **2016**, *12* (6), 555–559. https://doi.org/10.1038/nphys3671.

(42) Matsuno, J.; Ogawa, N.; Yasuda, K.; Kagawa, F.; Koshibae, W.; Nagaosa, N.; Tokura, Y.; Kawasaki, M. Interface-Driven Topological Hall Effect in SrRuO3-SrIrO3 Bilayer. *Sci. Adv.* **2016**, *2* (7), e1600304. https://doi.org/10.1126/sciadv.1600304.

(43) Gallagher, J. C.; Meng, K. Y.; Brangham, J. T.; Wang, H. L.; Esser, B. D.; McComb, D. W.; Yang, F. Y. Robust Zero-Field Skyrmion Formation in FeGe Epitaxial Thin Films. *Phys. Rev. Lett.* **2017**, *118*, 027201. https://doi.org/10.1103/PhysRevLett.118.027201.

(44) Liu, C.; Zang, Y.; Ruan, W.; Gong, Y.; He, K.; Ma, X.; Xue, Q. K.; Wang, Y. Dimensional Crossover-Induced Topological Hall Effect in a Magnetic Topological Insulator. *Phys. Rev. Lett.* **2017**, *119*, 176809. https://doi.org/10.1103/PhysRevLett.119.176809.

(45) Lindfors-Vrejoiu, I.; Ziese, M. Topological Hall Effect in Antiferromagnetically Coupled SrRuO3/La0.7Sr0.3MnO3 Epitaxial Heterostructures. *Phys. Status Solidi B* **2017**, *254* (5), 1600556. https://doi.org/10.1002/pssb.201600556.

(46) Liu, Z. H.; Zhang, Y. J.; Liu, G. D.; DIng, B.; Liu, E. K.; Jafri, H. M.; Hou, Z. P.;

Wang, W. H.; Ma, X. Q.; Wu, G. H. Transition from Anomalous Hall Effect to Topological Hall Effect in Hexagonal Non-Collinear Magnet Mn3Ga. *Sci. Rep.* **2017**, *7*, 515. https://doi.org/10.1038/s41598-017-00621-x.

(47) Ludbrook, B. M.; Dubuis, G.; Puichaud, A. H.; Ruck, B. J.; Granville, S. Nucleation and Annihilation of Skyrmions in Mn2CoAl Observed through the Topological Hall Effect. *Sci. Rep.* **2017**, *7*, 13620. https://doi.org/10.1038/s41598-017-13211-8.

(48) He, Q. L.; Yin, G.; Grutter, A. J.; Pan, L.; Che, X.; Yu, G.; Gilbert, D. A.; Disseler, S. M.; Liu, Y.; Shafer, P.; Zhang, B.; Wu, Y.; Kirby, B. J.; Arenholz, E.; Lake, R. K.; Han, X.; Wang, K. L. Exchange-Biasing Topological Charges by Antiferromagnetism. *Nat. Commun.* **2018**, *9*, 2767. https://doi.org/10.1038/s41467-018-05166-9.

(49) Ohuchi, Y.; Matsuno, J.; Ogawa, N.; Kozuka, Y.; Uchida, M.; Tokura, Y.; Kawasaki, M. Electric-Field Control of Anomalous and Topological Hall Effects in Oxide Bilayer Thin Films. *Nat. Commun.* **2018**, *9*, 213. https://doi.org/10.1038/s41467-017-02629-3.

(50) Meng, K. K.; Zhao, X. P.; Liu, P. F.; Liu, Q.; Wu, Y.; Li, Z. P.; Chen, J. K.; Miao, J.; Xu, X. G.; Zhao, J. H.; Jiang, Y. Robust Emergence of a Topological Hall Effect in MnGa/Heavy Metal Bilayers. *Phys. Rev. B* **2018**, *97*, 060407(R). https://doi.org/10.1103/PhysRevB.97.060407.

(51) Shao, Q.; Liu, Y.; Yu, G.; Kim, S. K.; Che, X.; Tang, C.; He, Q. L.; Tserkovnyak, Y.; Shi, J.; Wang, K. L. Topological Hall Effect at above Room Temperature in Heterostructures Composed of a Magnetic Insulator and a Heavy Metal. *Nat. Electron.* **2019**, *2* (5), 182–186. https://doi.org/10.1038/s41928-019-0246-x.

(52) Qin, Q.; Liu, L.; Lin, W.; Shu, X.; Xie, Q.; Lim, Z.; Li, C.; He, S.; Chow, G. M.; Chen, J. Emergence of Topological Hall Effect in a SrRuO3 Single Layer. *Adv. Mater.* **2019**, *31*, 1807008. https://doi.org/10.1002/adma.201807008.

(53) Lee, A. J.; Ahmed, A. S.; Flores, J.; Guo, S.; Wang, B.; Bagués, N.; McComb, D. W.; Yang, F. Probing the Source of the Interfacial Dzyaloshinskii-Moriya Interaction Responsible for the Topological Hall Effect in Metal/Tm3Fe5O12 Systems. *Phys. Rev. Lett.* **2020**, *124*, 107201. https://doi.org/10.1103/physrevlett.124.107201.

(54) Jiang, J.; Xiao, D.; Wang, F.; Shin, J. H.; Andreoli, D.; Zhang, J.; Xiao, R.; Zhao, Y. F.; Kayyalha, M.; Zhang, L.; Wang, K.; Zang, J.; Liu, C.; Samarth, N.; Chan, M. H. W.; Chang, C. Z. Concurrence of Quantum Anomalous Hall and Topological Hall Effects in Magnetic Topological Insulator Sandwich Heterostructures. *Nat. Mater.* **2020**, *19* (7), 732–737. https://doi.org/10.1038/s41563-020-0605-z.

(55) Sivakumar, P. K.; Göbel, B.; Lesne, E.; Markou, A.; Gidugu, J.; Taylor, J. M.; Deniz, H.; Jena, J.; Felser, C.; Mertig, I.; Parkin, S. S. P. Topological Hall Signatures of Two Chiral Spin Textures Hosted in a Single Tetragonal Inverse Heusler Thin Film. *ACS Nano* **2020**, *14* (10), 13463–13469. https://doi.org/10.1021/acsnano.0c05413.

(56) Yang, L.; Wysocki, L.; Schöpf, J.; Jin, L.; Kovács, A.; Gunkel, F.; Dittmann, R.;

Van Loosdrecht, P. H. M.; Lindfors-Vrejoiu, I. Origin of the Hump Anomalies in the Hall Resistance Loops of Ultrathin SrRuO3/SrIrO3 Multilayers. *Phys. Rev. Mater.* **2021**, *5*, 014403. https://doi.org/10.1103/PhysRevMaterials.5.014403.

(57) Zhang, X.; Ambhire, S. C.; Lu, Q.; Niu, W.; Cook, J.; Jiang, J. S.; Hong, D.; Alahmed, L.; He, L.; Zhang, R.; Xu, Y.; Zhang, S. S.-L.; Li, P.; Bian, G. Giant Topological Hall Effect in van Der Waals Heterostructures of CrTe2/Bi2Te3. *ACS Nano* **2021**, *15* (10), 15710-15719. https://doi.org/10.1021/acsnano.1c05519.

(58) Gerber, A. Interpretation of Experimental Evidence of the Topological Hall Effect. *Phys. Rev. B* **2018**, *98*, 214440. https://doi.org/10.1103/PhysRevB.98.214440.

(59) Liu, N.; Teng, J.; Li, Y. Two-Component Anomalous Hall Effect in a Magnetically Doped Topological Insulator. *Nat. Commun.* **2018**, *9*, 1282. https://doi.org/10.1038/s41467-018-03684-0.

(60) Fijalkowski, K. M.; Hartl, M.; Winnerlein, M.; Mandal, P.; Schreyeck, S.; Brunner, K.; Gould, C.; Molenkamp, L. W. Coexistence of Surface and Bulk Ferromagnetism Mimics Skyrmion Hall Effect in a Topological Insulator. *Phys. Rev. X* **2020**, *10*, 011012. https://doi.org/10.1103/PhysRevX.10.011012.

(61) Chen, P.; Zhang, Y.; Yao, Q.; Tian, F.; Li, L.; Qi, Z.; Liu, X.; Liao, L.; Song, C.; Wang, J.; Xia, J.; Li, G.; Burn, D. M.; Van Der Laan, G.; Hesjedal, T.; Zhang, S.; Kou, X. Tailoring the Hybrid Anomalous Hall Response in Engineered Magnetic Topological Insulator Heterostructures. *Nano Lett.* **2020**, *20* (3), 1731–1737. https://doi.org/10.1021/acs.nanolett.9b04932.

(62) Zhang, X.; Zhou, Y.; Song, K. M.; Park, T. E.; Xia, J.; Ezawa, M.; Liu, X.; Zhao, W.; Zhao, G.; Woo, S. Skyrmion-Electronics: Writing, Deleting, Reading and Processing Magnetic Skyrmions toward Spintronic Applications. *J. Phys. Condens. Matter* **2020**, *32* (14). https://doi.org/10.1088/1361-648X/ab5488.

(63) Otrokov, M. M.; Klimovskikh, I. I.; Bentmann, H.; Estyunin, D.; Zeugner, A.; Aliev, Z. S.; Gaß, S.; Wolter, A. U. B.; Koroleva, A. V.; Shikin, A. M.; Blanco-Rey, M.; Hoffmann, M.; Rusinov, I. P.; Vyazovskaya, A. Y.; Eremeev, S. V.; Koroteev, Y. M.; Kuznetsov, V. M.; Freyse, F.; Sánchez-Barriga, J.; Amiraslanov, I. R.; Babanly, M. B., *et al*. Prediction and Observation of an Antiferromagnetic Topological Insulator. *Nature* **2019**, *576* (7787), 416–422. https://doi.org/10.1038/s41586-019-1840-9.

(64) Li, H.; Liu, S.; Liu, C.; Zhang, J.; Xu, Y.; Yu, R.; Wu, Y.; Zhang, Y.; Fan, S. Antiferromagnetic Topological Insulator MnBi2Te4: Synthesis and Magnetic Properties. *Phys. Chem. Chem. Phys.* **2020**, *22* (2), 556–563. https://doi.org/10.1039/c9cp05634c.

(65) Zeugner, A.; Nietschke, F.; Wolter, A. U. B.; Gaß, S.; Vidal, R. C.; Peixoto, T. R. F.; Pohl, D.; Damm, C.; Lubk, A.; Hentrich, R.; Moser, S. K.; Fornari, C.; Min, C. H.; Schatz, S.; Kißner, K.; Ünzelmann, M.; Kaiser, M.; Scaravaggi, F.; Rellinghaus, B.; Nielsch, K., *et al*. Chemical Aspects of the Candidate Antiferromagnetic Topological Insulator MnBi2Te4. *Chem. Mater.* **2019**, *31* (8), 2795–2806. https://doi.org/10.1021/acs.chemmater.8b05017.

(66) Ding, L.; Hu, C.; Ye, F.; Feng, E.; Ni, N.; Cao, H. Crystal and Magnetic Structures

of Magnetic Topological Insulators MnBi2Te4 and MnBi4Te7. *Phys. Rev. B* **2020**, *101*, 020412(R). https://doi.org/10.1103/PhysRevB.101.020412.

(67) Yan, J. Q.; Zhang, Q.; Heitmann, T.; Huang, Z.; Chen, K. Y.; Cheng, J. G.; Wu, W.; Vaknin, D.; Sales, B. C.; McQueeney, R. J. Crystal Growth and Magnetic Structure of MnBi2Te4. *Phys. Rev. Mater.* **2019**, *3*, 064202. https://doi.org/10.1103/PhysRevMaterials.3.064202.

(68) Tai, L.; Li, J.; Chong, S. K.; Zhang, H.; Zhang, P.; Deng, P.; Eckberg, C.; Qiu, G.; Dai, B.; He, H.; Wu, D.; Xu, S.; Davydov, A. V; Wu, R.; Wang, K. L. Distinguishing Two-Component Anomalous Hall Effect from Topological Hall Effect in Magnetic Topological Insulator MnBi2Te4. *arXiv Prepr. arXiv2103.09878* **2021**, http://arxiv.org/abs/2103.09878.

TABLE OF CONTENTS GRAPHIC:

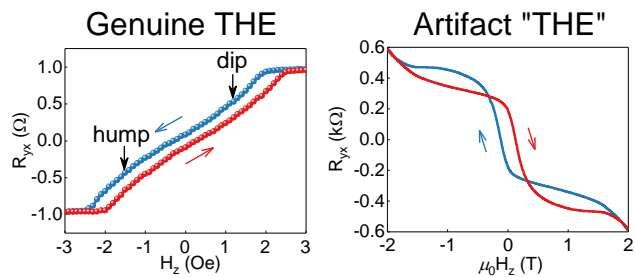

Supporting Information for

# "Distinguishing two-component anomalous Hall effect from topological Hall effect"

**Table of Contents**

# 1. Material characterizations of MnBi$_2$Te$_4$

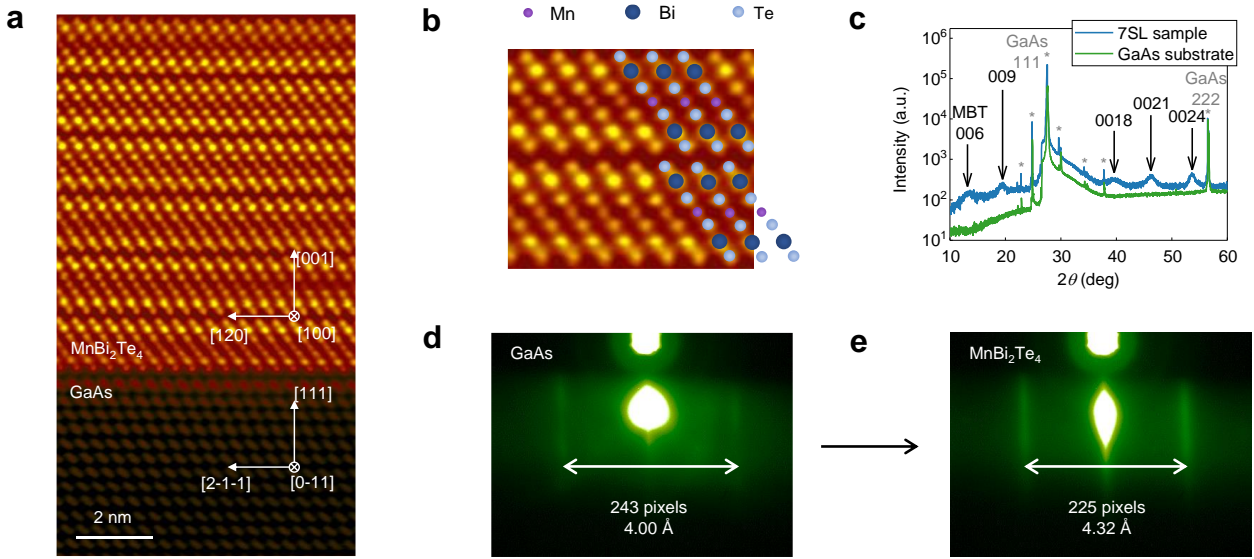

Figure S1. Material characterizations. (a) A cross-sectional HAADF-STEM image showing the epitaxial growth of MnBi$_2$Te$_4$ film on the GaAs substrate. The image was taken along the [100] zone-axis of the MnBi$_2$Te$_4$ crystal. (b) A local zoom-in view of (a) overlaid with an atomic model of MnBi$_2$Te$_4$. (c) The XRD patterns of the 7 SL sample and the GaAs substrate. Absence of all but 00$l$ ($l$=6,9,18,21,24) reflections corroborates STEM's data about the single-crystalline nature of the MnBi$_2$Te$_4$ film. The calculated c-lattice parameter is 40.876(3) Å. Note: the peaks from the substrate are marked with *. (d), (e) The RHEED patterns before and after the growth.

Among the various materials that exhibit the AHE or THE, magnetic topological insulators (MTI) are notable as they can realize the quantized version of the AHE, the quantum anomalous Hall effect (QAHE)[1–5] and can host exotic quasiparticle excitations that behave as axions [6–8] and Majorana fermions,[9,10] the latter of which has been shown to have potential for quantum computing. In addition, MTIs have very promising applications in energy-efficient magnetic memory devices because their unique topological surface state enables efficient charge to spin conversion and magnetization manipulation.[11,12] Recently, MnBi$_2$Te$_4$ has been discovered as an intrinsic MTI and is a layered tetradymite magnetic compound consisting of Te-Bi-Te-Mn-Te-Bi-Te septuple layers (SL) separated by van der Waals gaps (Fig. S1b). MnBi$_2$Te$_4$ is an A-type antiferromagnet (AFM), and it has a Néel temperature of 25.4 K. Each Mn$^{2+}$ ion has an out-of-plane magnetic moment of about 5$\mu_B$ (5 Bohr magnetons), which aligns parallel to each other within each SL, and anti-parallel between two neighboring SLs. Thin films of MnBi$_2$Te$_4$ can be made either through mechanical exfoliation of bulk crystals or molecular beam epitaxy (MBE), and intrinsically show a negative AHE (or QAHE) due to uncompensated magnetic moments. [4,7,8,13–24]

In this work, the thin films of MnBi$_2$Te$_4$ were grown on GaAs (111)B substrates by molecular beam epitaxy, and various methods are used to characterize the microstructural properties of the thin films.

High-angle annular dark field scanning transmission electron microscopy (HAADF-STEM) was carried out to study the atomic structures of the thin film. The HAADF-STEM characterization was performed on the 24 SL MnBi$_2$Te$_4$ sample with FEI Nova NanoLab 600 DualBeam (SEM/FIB). Initially, 0.5 μm thick Pt was deposited by electron beam-induced deposition on top of the thin film sample to protect its surface. After that, 1 μm Pt was deposited by ion beam-induced deposition. In the final step of preparation, the sample was cleaned with 2 kV Ga-ions using a low beam current of 29 pA and a small incident angle of 3 degrees to reduce Ga-ions damage. An FEI Titan 80-300 probe-corrected STEM/TEM microscope operating at 300 keV was employed to acquire atomic-resolution HAADF-STEM images.

Fig. S1a and b show the HAADF-STEM images with atomic resolution from a cross-sectional specimen. The images clearly show the characteristic septuple layer (SL) substructures of rhombic $MnBi_2Te_4$ epitaxially grown on cubic GaAs (111)B substrate with crystallographic relationships of $(001)_R \parallel (111)_C$ and $[100]_R \parallel [0\text{-}11]_C$. Heavier elements also reveal sharper contrasts (Bi>Te>Mn), which is consistent with the sequence of Te-Bi-Te-Mn-Te-Bi-Te in the SL structures. [15]

X-ray diffraction (XRD) was also performed to characterize the film's crystal structure. The XRD measurements were performed on the 7 SL $MnBi_2Te_4$ sample as well as the bare GaAs substrate using an X-ray powder diffractometer with Cu Kα radiation (Panalytical X'Pert Pro, λ= 1.5406 Å). The XRD pattern from the $MnBi_2Te_4$ sample in Fig. S1c shows five $00l$ ($l$=6,9,18,21,24) reflections from the $MnBi_2Te_4$ film and two high-intensity 111 and 222 peaks from GaAs, thus confirming the $(001)_R \parallel (111)_C$ epitaxial growth. The out-of-plane lattice constant $c$=40.876 ± 0.003 Å is calculated from the $00l$ XRD reflections (which comprises of 3 SLs, so that 1 SL= 13.63 Å), which agrees well with previous reports of the $MnBi_2Te_4$ lattice parameter values. [13,19] All the sharp, narrow peaks in the XRD pattern can be identified from the GaAs substrate and marked with *.

During growth, *in-situ* reflection high-energy electron diffraction (RHEED), as presented in Fig. S1d and e, showed sharp 1 × 1 diffraction streaks, indicating good epitaxial crystal quality and flat surface morphology. Because the spacing between the two first order diffraction streaks (*d*-spacing) is inversely proportional to the in-plane lattice constant, the in-plane lattice constant $a=4.32 \pm 0.02$ Å is extracted from the RHEED streaks, which aligns well with previous reports of the MnBi$_2$Te$_4$ lattice parameter values. [13,19]

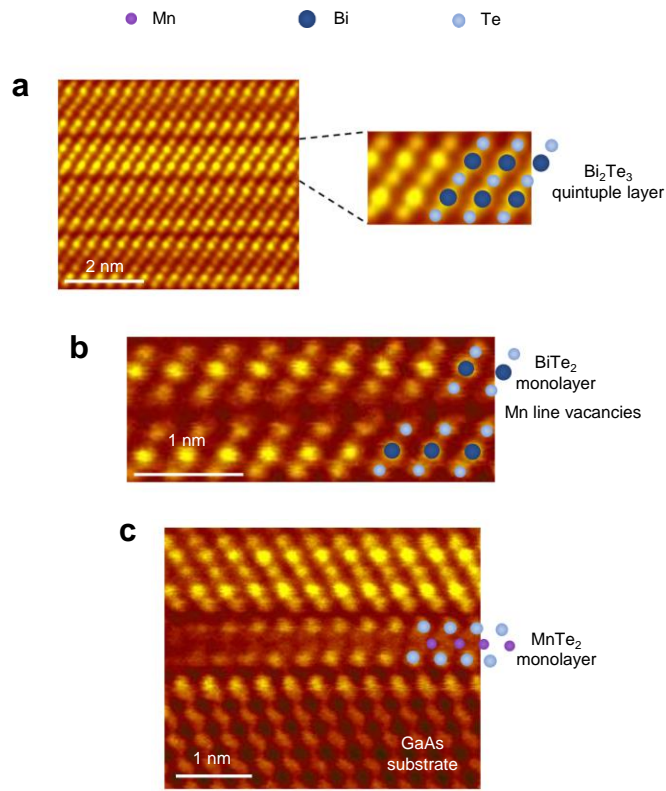

Figure S2. Additional cross-sectional HAADF-STEM images of the thin film, showing (a) a Bi$_2$Te$_3$ quintuple layer, (b) two BiTe$_2$ monolayers caused by Mn line vacancies in the MnBi$_2$Te$_4$ septuple layers and (c) a MnTe$_2$ monolayer on the surface of the thin film.

Notably, cross-sectional HAADF-STEM images of the thin film also reveal additional secondary phases, as shown in Fig. S2, including Bi$_2$Te$_3$ quintuple layers, BiTe$_2$ monolayers caused by Mn line vacancies in the MnBi$_2$Te$_4$ septuple layers, and a MnTe$_2$ monolayer on the surface. We will discuss the possible effects of these additional secondary phases in detail later.

## 2. The origins of two AHE components by DFT calculations

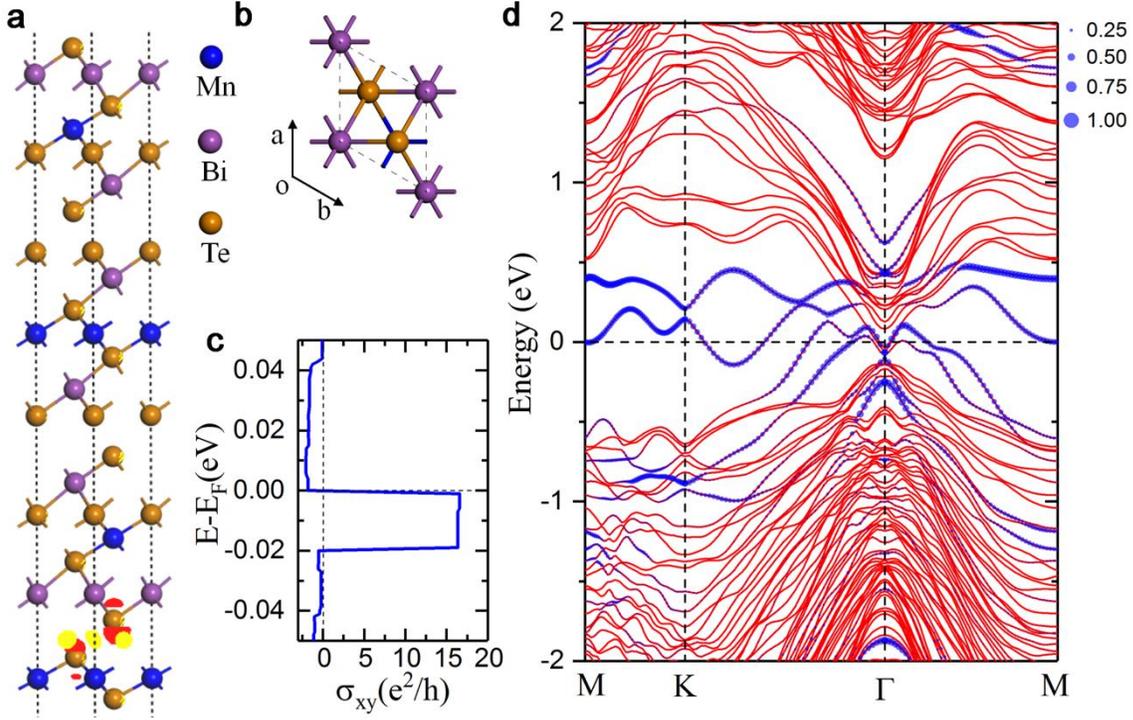

Figure S3. (a) The charge density redistribution of 3MBT-MT, (red and yellow colors indicate charge depletion and accumulation, and the isosurface is $3.0\times10^{-3}$ e/Å$^3$). (b) The top view of 3MBT-MT. (c) The Hall conductivity $\sigma_{xy}$ of 3MBT-MT as a function of Fermi level (d) The band structure of 3MBT-MT. (The blue circles represent the weight of MnTe$_2$ in the total band).

Now we are going to explore the origins of two AHE components, namely AH1 and AH2, by density functional theory (DFT) calculations. The negative AHE component AH1 is intrinsic to the magnetic topological insulator MnBi$_2$Te$_4$, which shows a negative AHE due to uncompensated magnetic moments, [4,15,25] while the positive AHE component AH2 is from a ferromagnetic MnTe$_2$ monolayer on the surface. [26,27]

All *ab initio* calculations in this work were carried out with the Vienna *ab-initio* simulation package (VASP) at the level of the spin-polarized generalized-gradient approximation (GGA).[28] The interaction between valence electrons and ionic cores was considered within the framework of the projector augmented wave (PAW) method.[29,30] The energy cutoff for the plane wave basis expansion was set to 500 eV. A Hubbard U = 4.0 eV was applied to the Manganese 3d orbitals. The vdW correction (DFT-D3) was included in all calculations.[31] All atoms were fully relaxed using the conjugated gradient method for the energy minimization until the force on each atom became smaller than 0.01 eV/Å, and $10^{-6}$ was chosen as the convergence criterion for the total energy for all DFT calculations.

Under cross-sectional HAADF-STEM in Fig. S2, we observed several additional secondary phases, including $Bi_2Te_3$ quintuple layers, $BiTe_2$ monolayers caused by Mn line vacancies in the $MnBi_2Te_4$ septuple layers, and a $MnTe_2$ monolayer on the surface. Among them, we believe that the interfacial $MnTe_2$ monolayer gives rise to the positive AH2. It is not likely for $Bi_2Te_3$ quintuple layers to give rise to the positive AH2, since they either do not contain magnetic elements and are non-magnetic, or contain Mn dopants and show a negative AHE. [32] $Bi_2Te_3$ can also work as non-magnetic buffer layers between $MnBi_2Te_4$ and give rise to weaker antiferromagnetic order or ferromagnetism in $MnBi_2Te_4/(Bi_2Te_3)_n$ superlattices, but the resulting AHE is still negative. [25,33–35] To the best of our knowledge, there is also no previous report of $BiTe_2$ monolayers giving rise to a positive AHE. It is most likely that the interfacial $MnTe_2$ monolayer gives rise to the positive AH2 since it is ferromagnetic and has perpendicular magnetic anisotropy (PMA) according to previous calculations. [26,27]

We performed density functional theory (DFT) calculations to further corroborate our hypothesis. Here in the calculations, we use a supercell with three septuple layers (SL) of MnBi$_2$Te$_4$, one monolayer of MnTe$_2$ (3MBT-MT) and a vacuum of 15Å thick along the surface normal as shown in Fig. S3a, and the following is the justification for this simplified model.

Considering the balance between calculation costs and simulation accuracy, only 3 SL MnBi$_2$Te$_4$ are included in this model. Our DFT calculations demonstrated that 3 SL MnBi$_2$Te$_4$ are enough to mimic the contributions of odd layers of MnBi$_2$Te$_4$ to the total Hall conductivity, because the 3 SL MnBi$_2$Te$_4$ film alone is a magnetic topological insulator with a gap of ~67 meV and a negative anomalous Hall conductivity ($-e^2/\hbar$) (see Fig. S4). This is in line with previous reports of the Hall conductivity of 5 or 7 SL MnBi$_2$Te$_4$.[4] On the other hand, an isolated MnTe$_2$ monolayer is metallic and has a large positive anomalous Hall conductivity, as shown in Fig. S5.

In this model, the optimized in-plane lattice constant and the interlayer distance between MnBi$_2$Te$_4$ and MnTe$_2$ are 4.34 Å and 2.42 Å, respectively, which agree well with the experimental results above. DFT calculations further show that the coupling between MnBi$_2$Te$_4$ and MnTe$_2$ at the interface is ferromagnetic with an exchange energy of ~15 meV per cell, and each Mn ion in MnBi$_2$Te$_4$ and MnTe$_2$ has a magnetic moment of $5\mu_B$ or $3\mu_B$, respectively. As shown in the charge density redistribution (see Fig. S3a), a small amount electron transfer from MnTe$_2$ to MnBi$_2$Te$_4$ occurs at the interface, and the Bader charge is ~0.03e per cell. This is also reflected in the band structure, where the Dirac cone of MnBi$_2$Te$_4$ shifts to below the Fermi level. Many interfacial states from a MnTe$_2$ monolayer cross the Fermi level, which offers a possibility for tuning the Hall conductivity of this system.

Quantitatively, the Hall conductivity $\sigma_{xy}$ as function of Fermi level was calculated from the Berry curvature over the Brillouin zone, which contains contributions from both MnBi$_2$Te$_4$ and the interfacial states from a MnTe$_2$ monolayer. As shown in Fig. S3c, the Hall conductivity can be tuned from positive (+16.37 $e^2/\hbar$) to negative (–2.07 $e^2/\hbar$) when the Fermi level is shifted up by 10 meV. Notably the interfacial states from a MnTe$_2$ monolayer play a dominant role in the large positive Hall conductivity. These results confirm that the negative AHE component AH1 is intrinsic to MnBi$_2$Te$_4$, and the positive AHE component AH2 originates from the interfacial MnTe$_2$ monolayer. These results also agree with the sign change of Hall resistance (as well as Hall conductivity) from positive to negative by increasing the gate bias, as shown in Fig. 6b and Fig. S9.

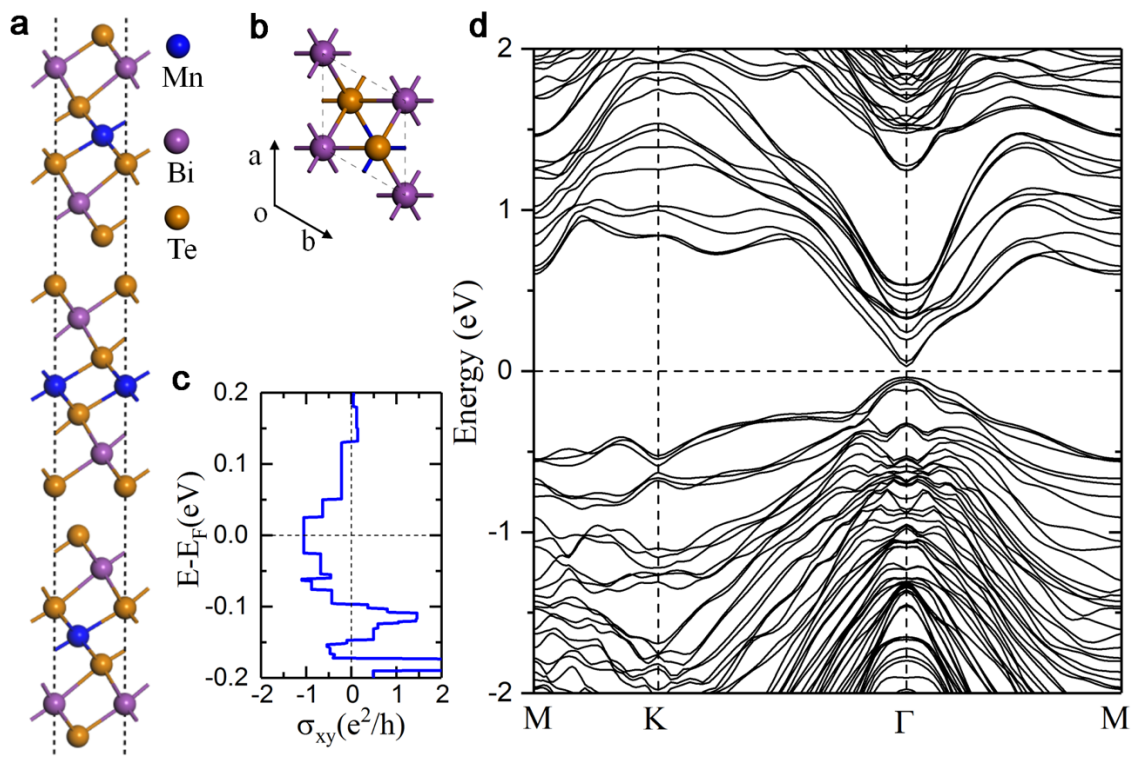

Figure S4. (a) (b) The side and top view of 3SL MnBi$_2$Te$_4$. (c) the Hall conductivity $\sigma_{xy}$ of 3SL MnBi$_2$Te$_4$ as a function of Fermi level (d) The band structure of 3SL MnBi$_2$Te$_4$.

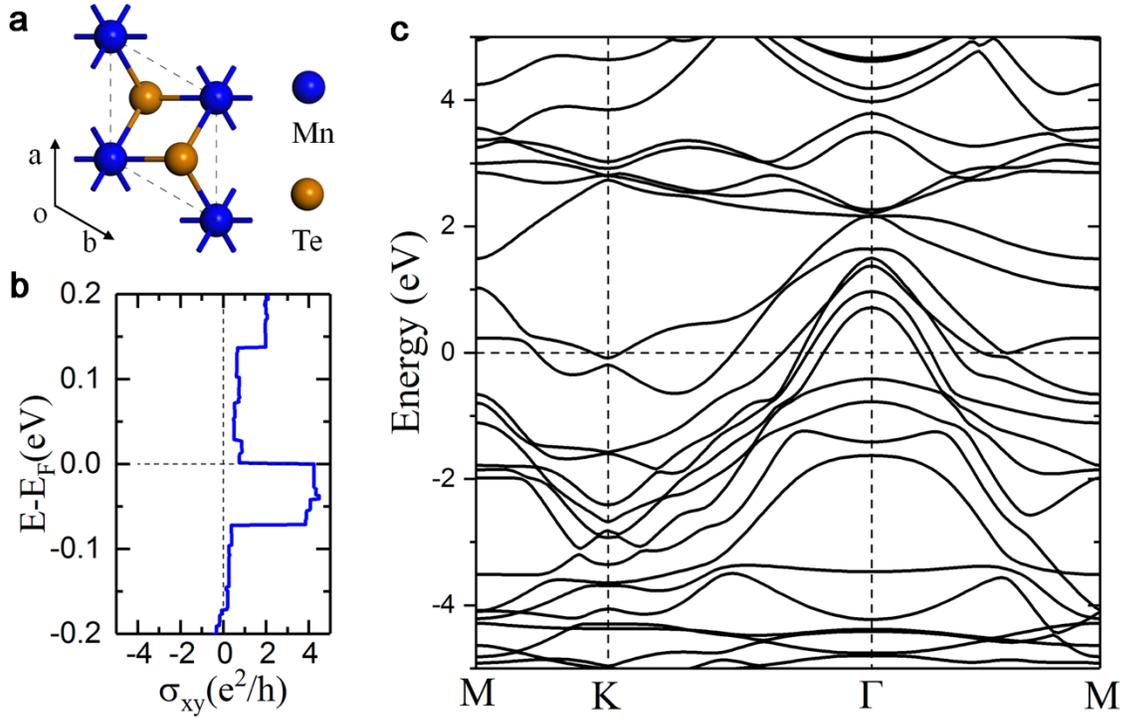

Figure S5. (a) The top view of monolayer MnTe$_2$. (b) The Hall conductivity σ$_{xy}$ of monolayer MnTe$_2$ as a function of Fermi level (c) The band structure of monolayer MnTe$_2$.

AH2 comes from the secondary phase of MnTe$_2$, and its gate dependence agrees with that of DFT calculations. Since there are two components of MnBi$_2$Te$_4$ and MnTe$_2$, their Hall conductivities add up together and form the total Hall conductivity. In the DFT calculations of Fig. S5, the Hall conductivity of MnTe$_2$ saturates and suddenly decreases with the increase of the Fermi level, while in Fig. S4, the Hall conductivity of the MnBi$_2$Te$_4$ follows an ambipolar response and does not change as much as MnTe$_2$ with Fermi level. As a result, MnTe$_2$'s percentage in the total Hall conductivity also first saturates and suddenly decreases rapidly with the increase of Fermi level, which corresponds to the increase of gate bias. This accounts for our experimental data in Fig. 6 that the AH2 component from MnTe$_2$ actually first saturates and then suddenly decreases with the increase of the gate bias.

## 3. The method of AHE decompositions

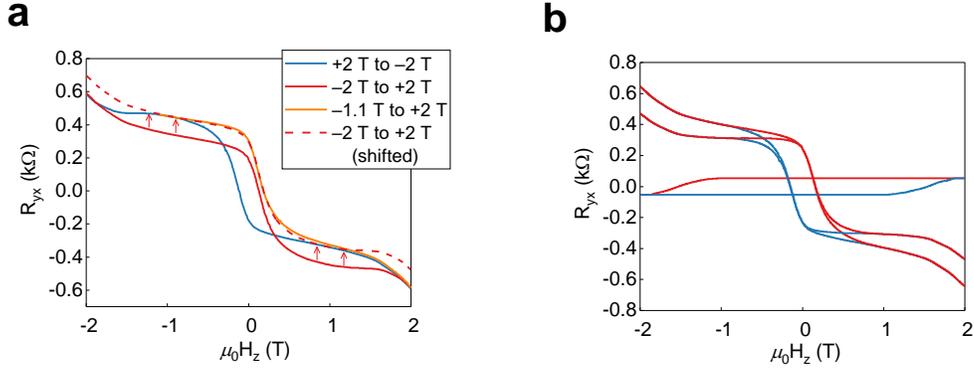

Figure S6. A schematic for the decomposition of the two-component AHE. (a) The full loop and the minor loop stopping at the field of -1.1 T, which shows no dip at positive fields (b) The decomposed data AH1 (with linear OHE), AH1 (with the OHE removed) and AH2.

The method of decomposition is as follows. Since within the field of $-1.1$ T, the AH2 remains unswitched, the minor loops contain only the AH1 component. The minor loop stopping at the field of $-1.1$ T is taken as the single AH1 to verify the decomposition. As shown in Fig. S6a, by shifting the forward sweeping curve from $-2$ T to $+2$ T upwards so that it becomes tangential to the backward sweeping curve from $+2$ T to $-2$ T, we can obtain a single AH1 loop that matches well with the minor loop. By subtracting the AH1 loop from the Hall signal, we can obtain the single AH2 loop, as shown in Fig. S6b. The full decompositions at various temperatures or gate biases are presented in Fig. S7 and Fig. S8.

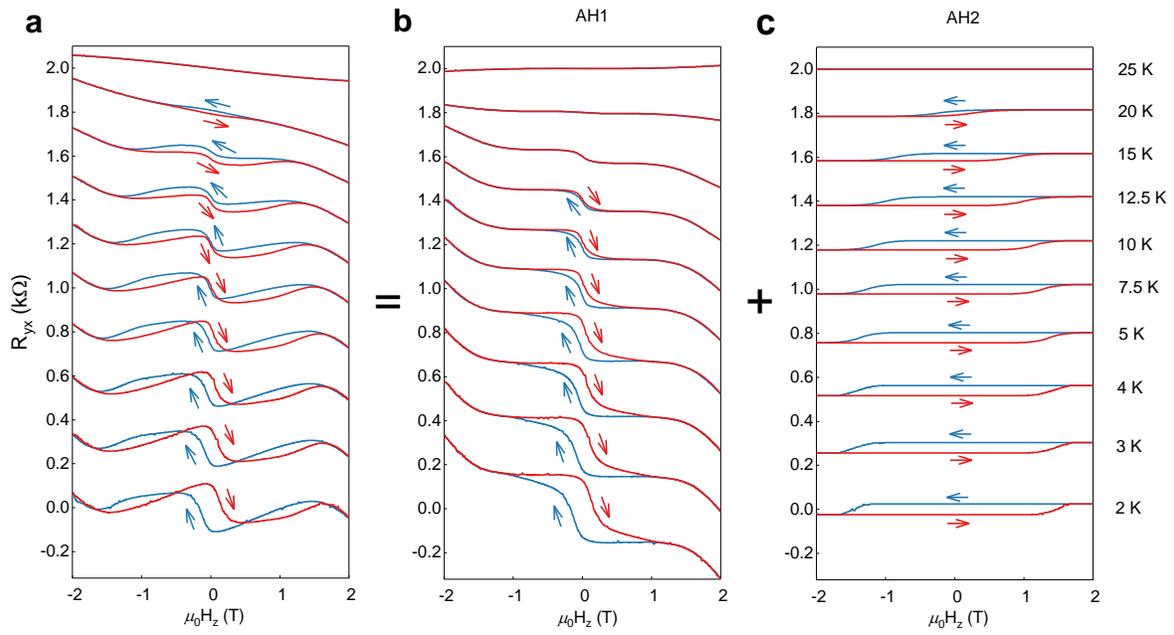

Figure S7. The temperature dependence of the two-component AHE in the 5 SL MnBi$_2$Te$_4$ sample. (a) The original data under various temperatures, with blue and red arrows indicating the direction of the field scan. (b), (c) The two AHEs that are obtained by decomposing the original data (AH1 and AH2).

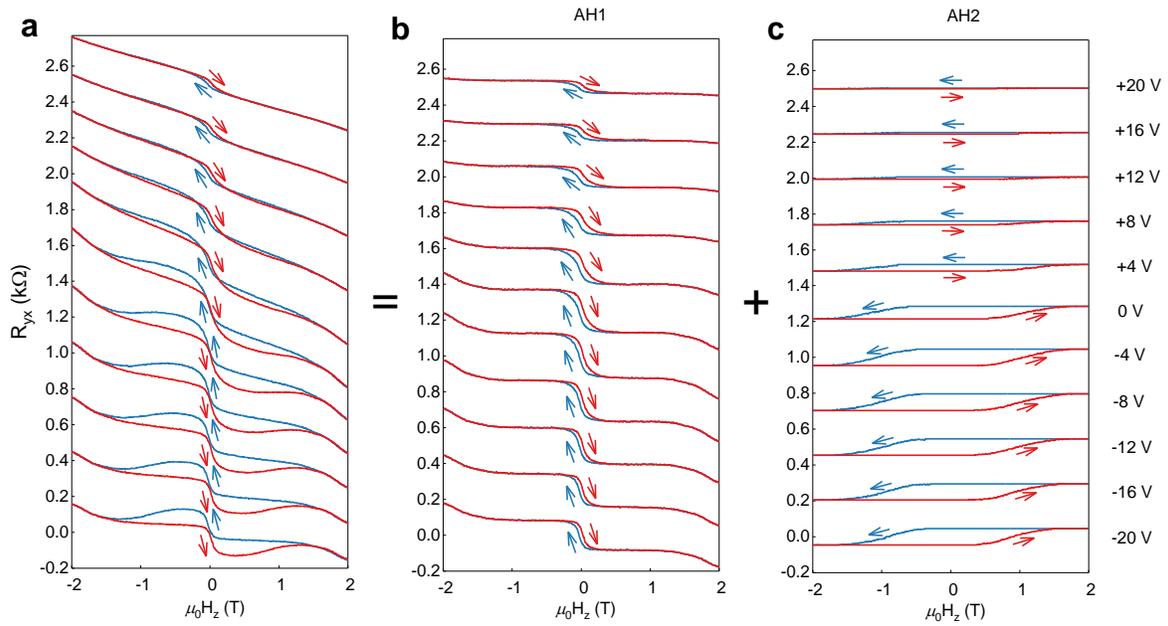

Figure S8. The gate dependence of the two-component AHE in the 7 SL MnBi$_2$Te$_4$ sample at T=10 K. (a) The original data under various gate biases, with blue and red arrows indicating the direction of the field scan. (b), (c) The two AHEs that are obtained by decomposing the original data (AH1 and AH2).

## 4. Gate dependence of Hall conductivity

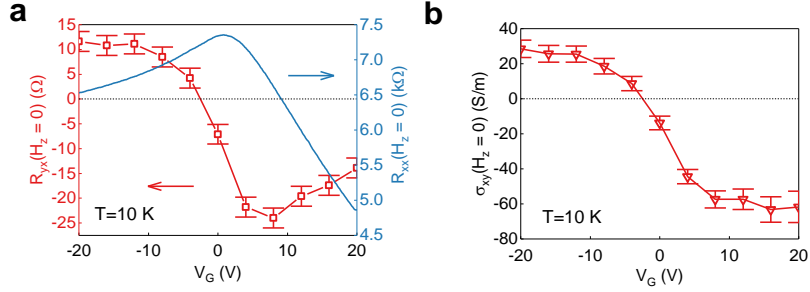

Figure S9. (a) The gate dependence of the zero-field Hall resistance $R_{yx}(H_z = 0)$ and longitudinal resistance $R_{xx}(H_z = 0)$ in the two-component AHE obtained from the 7 SL MnBi$_2$Te$_4$ sample (b) The gate dependence of the zero-field Hall conductivity, $\sigma_{xy}(H_z = 0)$.

In order to show the consistency between the experiments and the DFT calculations for the gate dependence of two-component AHE obtained from the 7 SL MnBi$_2$Te$_4$ sample, the gate dependence of zero-field Hall resistance $R_{yx}(H_z = 0)$ and longitudinal resistance $R_{xx}(H_z = 0)$ is presented in Fig. S9a, and the gate dependence of zero-field Hall conductivity, $\sigma_{xy}(H_z = 0)$ is presented in Fig. S9b, which is converted by

$$\sigma_{xy} = \frac{\rho_{yx}}{\rho_{xx}^2 + \rho_{yx}^2}$$

The resistivities $\rho_i$ are obtained from the resistances by $R_i = \rho_i \, l/S = \rho_i \, l/(dt)$, where $i = xx$ or $yx$, $l = 40$ μm is the length, $d = 40$ μm is the width, and $t = 9.54$ nm is the thickness of the Hall bar. Our experiments in Fig. 6b and Fig. S9 demonstrate that the Hall resistance, as well as Hall conductivity changes sign from positive to negative by increasing the gate bias, which increases the Fermi level of the sample. In comparison, our DFT calculations in Fig. S3 also show that the Hall conductivity changes sign from positive to negative with the increase of Fermi level, capturing the sign change of Hall data with gate.

## 5. Thickness dependence

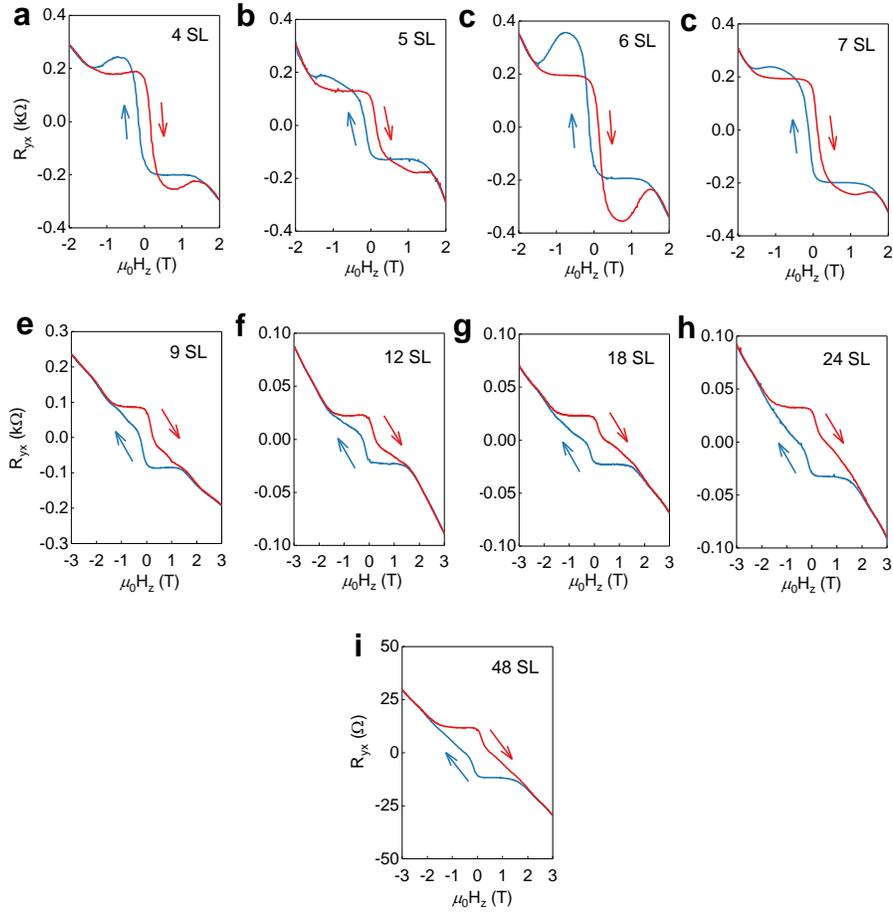

Figure S10. The thickness dependence of the AHE in the MnBi$_2$Te$_4$ samples. (a)-(i) The Hall resistance of samples with various thicknesses as a function of out-of-plane magnetic field $H_z$ at T=2 K, with blue and red arrows indicating the direction of the field scan. The linear ordinary Hall backgrounds are removed.

| Thickness (SL) | 4 | 5 | 6 | 7 | 9 | 12 | 18 | 24 | 48 |
|---|---|---|---|---|---|---|---|---|---|
| Have humps and dips | Yes | Yes | Yes | Yes | No | No | No | No | No |
| $H_C$ (T) | 0.16 | 0.15 | 0.13 | 0.12 | 0.27 | 0.25 | 0.44 | 0.55 | 0.51 |

Table S1. The existence of humps and dips in the AHE loops and the coercivity ($H_C$) of the AHE for samples with various thicknesses.

The thickness dependence of the AHE loops in the MnBi$_2$Te$_4$ samples is presented in Fig. S10. In order to extract the coercivity from the intersection of the AHE loops and the *x*-axis, the linear ordinary Hall backgrounds have been subtracted around the saturation field of ~+0.7 T (when sweeping from +2 T to –2 T). All the samples show a negative AHE under the temperature at T=2 K due to uncompensated magnetic moments of MnBi$_2$Te$_4$ (the even layers still have uncompensated magnetic moments due to variations of thickness over different areas, so the thickness is only taken as an average thickness) However, the humps and dips in the AHE only appear between 4 to 7 SL and vanish for the thickness of 9 SL and beyond. Vanishing humps and dips with increasing thickness are consistent with our conclusion that the interfacial MnTe$_2$ monolayer gives rise to the positive AH2. As the transport becomes dominated by the bulk channels in thicker layers, the AH2 from the surface is no longer distinguishable. As the thickness further increases, the coercivities also tend to increase to a large extent, although not monotonously, as are listed in Table S1.

## 6. Reproducibility and controllability

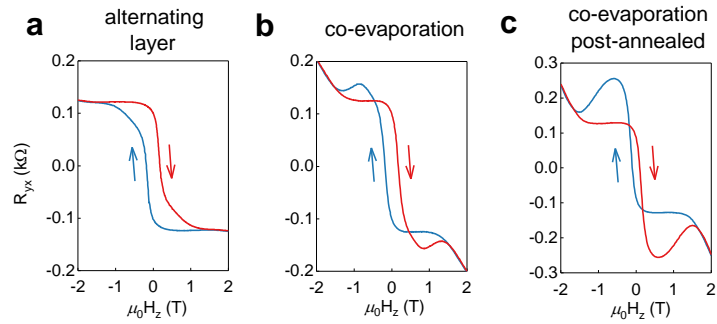

Figure S11. The Hall resistance of the three 6 SL MnBi$_2$Te$_4$ samples with different growth methods and the same medium Mn concentration as a function of out-of-plane magnetic field $H_z$ at T=2 K, with blue and red arrows indicating the direction of the field scan. The linear ordinary Hall background is removed.

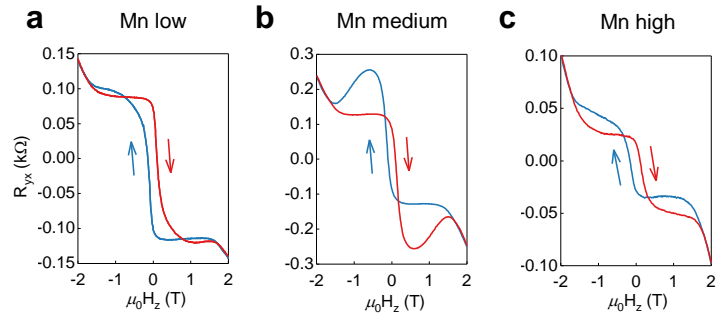

Figure S12. The Hall resistance of three 6 SL MnBi$_2$Te$_4$ samples grown by the same co-evaporation + post-annealing method with various Mn concentrations as a function of out-of-plane magnetic field $H_z$ at T=2 K, with blue and red arrows indicating the direction of the field scan. The linear ordinary Hall background is removed.

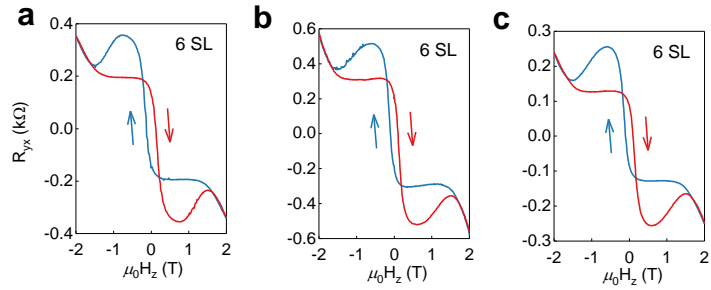

Figure S13. The Hall resistance of three different 6 SL MnBi$_2$Te$_4$ samples grown with the same co-evaporation + post-annealing method and the same medium Mn concentration as a function of out-of-plane magnetic field $H_z$ at T=2 K, with blue and red arrows indicating the direction of the field scan. The linear ordinary Hall background is removed.

To show that the humps and dips in the Hall signal caused by the additional positive AHE component AH2 are both controllable and repeatable, we present a systematic study of growth methods of MnBi$_2$Te$_4$.

The humps and dips in the Hall signal can be induced by co-evaporation and post-annealing, as shown by three 6 SL samples with different growth methods and the same medium Mn concentration in Fig. S11. Here co-evaporation means that Mn, Bi and Te were evaporated simultaneously during the growth, and alternating layer means depositing 1 quintuple layer (QL) Bi$_2$Te$_3$ and 1 bilayer (BL) MnTe alternatively (thus evaporating Bi and Mn alternatively) to automatically form 1 SL MnBi$_2$Te$_4$, as demonstrated by Gong *et al*. [15] Post-annealing means after the deposition, the film was post-annealed in a Te-rich environment at 290 °C for 2 minutes to improve crystallinity.

We found that the growth method of alternating layer gave rise to a single negative AHE without humps and dips, which is consistent with Gong *et al*. [15] However, as the growth method was changed to co-evaporation, humps and dips in the Hall signal emerged. Moreover, as post-annealing was further introduced, the relative amplitude of the humps and dips increased dramatically. This can be explained by the fact that the secondary phase of $MnTe_2$ on the surface can be induced effectively by means of co-evaporation and post-annealing, thus contributing to an extra positive AHE component AH2.

The relative amplitude of the humps and dips also increases with Mn concentration in the sample, as indicated by three 6 SL samples grown by the same co-evaporation + post-annealing method with various Mn concentrations in Fig. S11. Here low, medium and high Mn concentrations mean that the standard Knudsen cell for Mn was kept at 730, 750, and 770°C during growth, respectively, and a higher temperature means higher Mn vapor pressure and thus higher Mn concentration in the sample. A low Mn concentration gave rise to relatively smaller humps and dips because less secondary phase of $MnTe_2$ (and more secondary phase of Mn-doped $Bi_2Te_3$) was present. As Mn concentration increased to medium and more secondary phase of $MnTe_2$ was formed, the relative amplitude of the humps and dips increased significantly. When Mn concentration further increased to high, the AHE shrank dramatically, thus making the relative contribution of humps and dips even larger, as even more metallic $MnTe_2$ was present and AH2 caused by $MnTe_2$ dominated the conduction channel over the negative AH1 from $MnBi_2Te_4$ itself.

The humps and dips in the Hall signal are also highly repeatable, as confirmed by three different 6 SL samples grown with the same co-evaporation + post-annealing method and the same medium Mn concentration in Fig. S12. The Hall signals reveal a similar pattern of humps and dips, despite some difference in the amplitude caused by some geometric difference in the soldering process or variations in the growth. This means that the relative contribution of AH2 from $MnTe_2$ is similar with the same growth method.

**References**

(1) Chang, C. Z.; Zhang, J.; Feng, X.; Shen, J.; Zhang, Z.; Guo, M.; Li, K.; Ou, Y.; Wei, P.; Wang, L. L.; Ji, Z. Q.; Feng, Y.; Ji, S.; Chen, X.; Jia, J.; Dai, X.; Fang, Z.; Zhang, S. C.; He, K.; Wang, Y., *et al*. Experimental Observation of the Quantum Anomalous Hall Effect in a Magnetic Topological Insulator. *Science* **2013**, *340* (6129), 167–170. https://doi.org/10.1126/science.1234414.

(2) Kou, X.; Pan, L.; Wang, J.; Fan, Y.; Choi, E. S.; Lee, W. L.; Nie, T.; Murata, K.; Shao, Q.; Zhang, S. C.; Wang, K. L. Metal-to-Insulator Switching in Quantum Anomalous Hall States. *Nat. Commun.* **2015**, *6*, 8474. https://doi.org/10.1038/ncomms9474.

(3) Kou, X.; Guo, S. T.; Fan, Y.; Pan, L.; Lang, M.; Jiang, Y.; Shao, Q.; Nie, T.; Murata, K.; Tang, J.; Wang, Y.; He, L.; Lee, T. K.; Lee, W. L.; Wang, K. L. Scale-Invariant Quantum Anomalous Hall Effect in Magnetic Topological Insulators beyond the Two-Dimensional Limit. *Phys. Rev. Lett.* **2014**, *113*, 137201. https://doi.org/10.1103/PhysRevLett.113.137201.

(4) Deng, Y.; Yu, Y.; Shi, M. Z.; Guo, Z.; Xu, Z.; Wang, J.; Chen, X. H.; Zhang, Y. Quantum Anomalous Hall Effect in Intrinsic Magnetic Topological Insulator MnBi2Te4. *Science* **2020**, *367* (6480), 895–900. https://doi.org/10.1126/science.aax8156.

(5) Checkelsky, J. G.; Yoshimi, R.; Tsukazaki, A.; Takahashi, K. S.; Kozuka, Y.; Falson, J.; Kawasaki, M.; Tokura, Y. Trajectory of the Anomalous Hall Effect towards the Quantized State in a Ferromagnetic Topological Insulator. *Nat. Phys.* **2014**, *10* (10), 731–736. https://doi.org/10.1038/nphys3053.

(6) Xiao, D.; Jiang, J.; Shin, J. H.; Wang, W.; Wang, F.; Zhao, Y. F.; Liu, C.; Wu, W.; Chan, M. H. W.; Samarth, N.; Chang, C. Z. Realization of the Axion Insulator State in Quantum Anomalous Hall Sandwich Heterostructures. *Phys. Rev. Lett.* **2018**, *120* (5), 56801. https://doi.org/10.1103/PhysRevLett.120.056801.

(7) Liu, C.; Wang, Y.; Li, H.; Wu, Y.; Li, Y.; Li, J.; He, K.; Xu, Y.; Zhang, J.; Wang, Y. Robust Axion Insulator and Chern Insulator Phases in a Two-Dimensional Antiferromagnetic Topological Insulator. *Nat. Mater.* **2020**, *19*, 522–527.

https://doi.org/10.1038/s41563-019-0573-3.

(8) Zhang, D.; Shi, M.; Zhu, T.; Xing, D.; Zhang, H.; Wang, J. Topological Axion States in the Magnetic Insulator MnBi2Te4 with the Quantized Magnetoelectric Effect. *Phys. Rev. Lett.* **2019**, *122* (20), 206401. https://doi.org/10.1103/PhysRevLett.122.206401.

(9) He, Q. L.; Pan, L.; Stern, A. L.; Burks, E. C.; Che, X.; Yin, G.; Wang, J.; Lian, B.; Zhou, Q.; Choi, E. S.; Murata, K.; Kou, X.; Chen, Z.; Nie, T.; Shao, Q.; Fan, Y.; Zhang, S. C.; Liu, K.; Xia, J.; Wang, K. L. Chiral Majorana Fermion Modes in a Quantum Anomalous Hall Insulator–superconductor Structure. *Science* **2017**, *357* (6348), 294–299. https://doi.org/10.1126/science.aag2792.

(10) Peng, Y.; Xu, Y. Proximity-Induced Majorana Hinge Modes in Antiferromagnetic Topological Insulators. *Phys. Rev. B* **2019**, *99*, 195431. https://doi.org/10.1103/PhysRevB.99.195431.

(11) Che, X.; Pan, Q.; Vareskic, B.; Zou, J.; Pan, L.; Zhang, P.; Yin, G.; Wu, H.; Shao, Q.; Deng, P.; Wang, K. L. Strongly Surface State Carrier-Dependent Spin–Orbit Torque in Magnetic Topological Insulators. *Adv. Mater.* **2020**, *32*, 1907661. https://doi.org/10.1002/adma.201907661.

(12) Fan, Y.; Upadhyaya, P.; Kou, X.; Lang, M.; Takei, S.; Wang, Z.; Tang, J.; He, L.; Chang, L. Te; Montazeri, M.; Yu, G.; Jiang, W.; Nie, T.; Schwartz, R. N.; Tserkovnyak, Y.; Wang, K. L. Magnetization Switching through Giant Spin-Orbit Torque in a Magnetically Doped Topological Insulator Heterostructure. *Nat. Mater.* **2014**, *13* (7), 699–704. https://doi.org/10.1038/nmat3973.

(13) Lee, D. S.; Kim, T. H.; Park, C. H.; Chung, C. Y.; Lim, Y. S.; Seo, W. S.; Park, H. H. Crystal Structure, Properties and Nanostructuring of a New Layered Chalcogenide Semiconductor, Bi2MnTe4. *CrystEngComm* **2013**, *15* (27), 5532–5538. https://doi.org/10.1039/c3ce40643a.

(14) Otrokov, M. M.; Klimovskikh, I. I.; Bentmann, H.; Estyunin, D.; Zeugner, A.; Aliev, Z. S.; Gaß, S.; Wolter, A. U. B.; Koroleva, A. V.; Shikin, A. M.; Blanco-Rey, M.; Hoffmann, M.; Rusinov, I. P.; Vyazovskaya, A. Y.; Eremeev, S. V.; Koroteev, Y. M.; Kuznetsov, V. M.; Freyse, F.; Sánchez-Barriga, J.; Amiraslanov, I. R.; Babanly, M. B., *et al*. Prediction and Observation of an Antiferromagnetic Topological Insulator. *Nature* **2019**, *576* (7787), 416–422. https://doi.org/10.1038/s41586-019-1840-9.

(15) Gong, Y.; Guo, J.; Li, J.; Zhu, K.; Liao, M.; Liu, X.; Zhang, Q.; Gu, L.; Tang, L.; Feng, X.; Zhang, D.; Li, W.; Song, C.; Wang, L.; Yu, P.; Chen, X.; Wang, Y.; Yao, H.; Duan, W.; Xu, Y., *et al*. Experimental Realization of an Intrinsic Magnetic Topological Insulator. *Chinese Phys. Lett.* **2019**, *36* (7), 076801. https://doi.org/10.1088/0256-307X/36/7/076801.

(16) Hirahara, T.; Eremeev, S. V.; Shirasawa, T.; Okuyama, Y.; Kubo, T.; Nakanishi, R.; Akiyama, R.; Takayama, A.; Hajiri, T.; Ideta, S. I.; Matsunami, M.; Sumida, K.; Miyamoto, K.; Takagi, Y.; Tanaka, K.; Okuda, T.; Yokoyama, T.; Kimura, S.

I.; Hasegawa, S.; Chulkov, E. V. Large-Gap Magnetic Topological Heterostructure Formed by Subsurface Incorporation of a Ferromagnetic Layer. *Nano Lett.* **2017**, *17* (6), 3493–3500. https://doi.org/10.1021/acs.nanolett.7b00560.

(17) Rienks, E. D. L.; Wimmer, S.; Sánchez-Barriga, J.; Caha, O.; Mandal, P. S.; Růžička, J.; Ney, A.; Steiner, H.; Volobuev, V. V.; Groiss, H.; Albu, M.; Kothleitner, G.; Michalička, J.; Khan, S. A.; Minár, J.; Ebert, H.; Bauer, G.; Freyse, F.; Varykhalov, A.; Rader, O.; Springholz, G. Large Magnetic Gap at the Dirac Point in Bi2Te3/MnBi2Te4 Heterostructures. *Nature* **2019**, *576* (7787), 423–428. https://doi.org/10.1038/s41586-019-1826-7.

(18) Ge, J.; Liu, Y.; Li, J.; Li, H.; Luo, T.; Wu, Y.; Xu, Y.; Wang, J. High-Chern-Number and High-Temperature Quantum Hall Effect without Landau Levels. *Natl. Sci. Rev.* **2020**, *7* (8), 1280–1287. https://doi.org/10.1093/nsr/nwaa089.

(19) Yan, J. Q.; Zhang, Q.; Heitmann, T.; Huang, Z.; Chen, K. Y.; Cheng, J. G.; Wu, W.; Vaknin, D.; Sales, B. C.; McQueeney, R. J. Crystal Growth and Magnetic Structure of MnBi2Te4. *Phys. Rev. Mater.* **2019**, *3*, 064202. https://doi.org/10.1103/PhysRevMaterials.3.064202.

(20) Li, H.; Liu, S.; Liu, C.; Zhang, J.; Xu, Y.; Yu, R.; Wu, Y.; Zhang, Y.; Fan, S. Antiferromagnetic Topological Insulator MnBi2Te4: Synthesis and Magnetic Properties. *Phys. Chem. Chem. Phys.* **2020**, *22* (2), 556–563. https://doi.org/10.1039/c9cp05634c.

(21) Zeugner, A.; Nietschke, F.; Wolter, A. U. B.; Gaß, S.; Vidal, R. C.; Peixoto, T. R. F.; Pohl, D.; Damm, C.; Lubk, A.; Hentrich, R.; Moser, S. K.; Fornari, C.; Min, C. H.; Schatz, S.; Kißner, K.; Ünzelmann, M.; Kaiser, M.; Scaravaggi, F.; Rellinghaus, B.; Nielsch, K., *et al.* Chemical Aspects of the Candidate Antiferromagnetic Topological Insulator MnBi2Te4. *Chem. Mater.* **2019**, *31* (8), 2795–2806. https://doi.org/10.1021/acs.chemmater.8b05017.

(22) Ding, L.; Hu, C.; Ye, F.; Feng, E.; Ni, N.; Cao, H. Crystal and Magnetic Structures of Magnetic Topological Insulators MnBi2Te4 and MnBi4Te7. *Phys. Rev. B* **2020**, *101*, 020412(R). https://doi.org/10.1103/PhysRevB.101.020412.

(23) Li, J.; Li, Y.; Du, S.; Wang, Z.; Gu, B. L.; Zhang, S. C.; He, K.; Duan, W.; Xu, Y. Intrinsic Magnetic Topological Insulators in van Der Waals Layered MnBi2Te4-Family Materials. *Sci. Adv.* **2019**, *5* (6), eaaw5685. https://doi.org/10.1126/sciadv.aaw5685.

(24) Otrokov, M. M.; Rusinov, I. P.; Blanco-Rey, M.; Hoffmann, M.; Vyazovskaya, A. Y.; Eremeev, S. V.; Ernst, A.; Echenique, P. M.; Arnau, A.; Chulkov, E. V. Unique Thickness-Dependent Properties of the van Der Waals Interlayer Antiferromagnet MnBi2Te4 Films. *Phys. Rev. Lett.* **2019**, *122* (10), 107202. https://doi.org/10.1103/PhysRevLett.122.107202.

(25) Deng, H.; Chen, Z.; Wolos, A.; Konczykowski, M.; Sobczak, K.; Sitnicka, J.; Fedorchenko, I. V.; Borysiuk, J.; Heider, T.; Plucinski, L.; Park, K.; Georgescu, A. B.; Cano, J.; Krusin-Elbaum, L. Observation of High-Temperature Quantum

Anomalous Hall Regime in Intrinsic MnBi2Te4/Bi2Te3 Superlattice. *Nat. Phys.* **2020**, *17* (1), 36–42. https://doi.org/10.1038/s41567-020-0998-2.

(26) Chen, W.; Zhang, J. M.; Nie, Y. Z.; Xia, Q. L.; Guo, G. H. Electronic Structure and Magnetism of MTe2 (M = Ti, V, Cr, Mn, Fe, Co and Ni) Monolayers. *J. Magn. Magn. Mater.* **2020**, *508*, 166878. https://doi.org/10.1016/j.jmmm.2020.166878.

(27) Chen, W.; Zhang, J. M.; Nie, Y. Z.; Xia, Q. L.; Guo, G. H. Tuning Magnetic Properties of Single-Layer MnTe2 via Strain Engineering. *J. Phys. Chem. Solids* **2020**, *143*, 109489. https://doi.org/10.1016/j.jpcs.2020.109489.

(28) Perdew, J. P.; Burke, K.; Ernzerhof, M. Generalized Gradient Approximation Made Simple. *Phys. Rev. Lett.* **1996**, *77* (18), 3865–3868. https://doi.org/10.1103/PhysRevLett.77.3865.

(29) Blöchl, P. E. Projector Augmented-Wave Method. *Phys. Rev. B* **1994**, *50* (24), 17953–17979. https://doi.org/10.1103/PhysRevB.50.17953.

(30) Kresse, G.; Joubert, D. From Ultrasoft Pseudopotentials to the Projector Augmented-Wave Method. *Phys. Rev. B* **1999**, *59* (3), 1758–1775. https://doi.org/10.1103/PhysRevB.59.1758.

(31) Grimme, S.; Antony, J.; Ehrlich, S.; Krieg, H. A Consistent and Accurate Ab Initio Parametrization of Density Functional Dispersion Correction (DFT-D) for the 94 Elements H-Pu. *J. Chem. Phys.* **2010**, *132*, 154104. https://doi.org/10.1063/1.3382344.

(32) Checkelsky, J. G.; Ye, J.; Onose, Y.; Iwasa, Y.; Tokura, Y. Dirac-Fermion-Mediated Ferromagnetism in a Topological Insulator. *Nat. Phys.* **2012**, *8* (10), 729–733. https://doi.org/10.1038/nphys2388.

(33) Wu, J.; Liu, F.; Sasase, M.; Ienaga, K.; Obata, Y.; Yukawa, R.; Horiba, K.; Kumigashira, H.; Okuma, S.; Inoshita, T.; Hosono, H. Natural van Der Waals Heterostructural Single Crystals with Both Magnetic and Topological Properties. *Sci. Adv.* **2019**, *5*, eaax9989. https://doi.org/10.1126/sciadv.aax9989.

(34) Tian, S.; Li, H.; Wang, Z. Magnetic Topological Insulator MnBi6Te10 with a Zero-Field Ferromagnetic State and Gapped Dirac Surface States. *Phys. Rev. B* **2020**, *102* (3), 35144. https://doi.org/10.1103/PhysRevB.102.035144.

(35) Hu, C.; Ding, L.; Gordon, K. N.; Ghosh, B.; Tien, H. J.; Li, H.; Garrison Linn, A.; Lien, S. W.; Huang, C. Y.; Mackey, S.; Liu, J.; Sreenivasa Reddy, P. V.; Singh, B.; Agarwal, A.; Bansil, A.; Song, M.; Li, D.; Xu, S. Y.; Lin, H.; Cao, H.; Chang, T. R.; Dessau, D.; Ni, N. Realization of an Intrinsic Ferromagnetic Topological State in MnBi8Te13. *Sci. Adv.* **2020**, *6*, eaba4275. https://doi.org/10.1126/sciadv.aba4275.